# Lattice parameters and stability of the spinel compounds in relation to the ionic radii and electronegativities of constituting chemical elements

**Mikhail G. Brik[1,2,*], Andrzej Suchocki[3,4], Agata Kamińska[3]**

[1] *College of Mathematics and Physics, Chongqing University of Posts and Telecommunications, Chongqing 400065, P.R. China*
[2] *Institute of Physics, University of Tartu, Riia 142, Tartu 51014, Estonia*
[3] *Institute of Physics, Polish Academy of Sciences, al. Lotników 32/46, 02-668 Warsaw, Poland*
[4] *Institute of Physics, Kazimierz Wielki University, Weyssenhoffa 11, 85-072 Bydgoszcz, Poland*

[*] Corresponding author. E-mail: brik@fi.tartu.ee Phone: +372 7374751; Fax: +372 738 3033

## Abstract

A thorough consideration of the relation between the lattice parameters of 185 binary and ternary spinel compounds, on one side, and ionic radii and electronegativities of the constituting ions, on the other side, allowed for establishing a simple empirical model and finding its linear equation, which links together the above-mentioned quantities. The derived equation gives good agreement between the experimental and modeled values of the lattice parameters in the considered group of spinels, with an average relative error of about 1% only. The proposed model was improved further by separate consideration of several groups of spinels, depending on the nature of the anion (oxygen, sulfur, selenium/tellurium, nitrogen). The developed approach can be efficiently used for prediction of lattice constants for new isostructural materials. In particular, the lattice constants of new hypothetic spinels $ZnRE_2O_4$, $CdRE_2$S4, $CdRE_2Se_4$ (RE=rare earth elements) are predicted in the present paper. In addition, the upper and lower limits for the variation of the ionic radii, electronegativities and certain their combinations were established, which can be considered as stability criteria for the spinel compounds. The findings of the present paper offer a systematic overview of the structural properties of spinels and can serve as helpful guides for synthesis of new spinel compounds.

## 1. Introduction

Crystal structure of any crystalline material can be described in a unique way by giving the values of the unit cell parameters (lattice constants – LCs – along each of the crystallographic axes and angles between these axes), atomic positions expressed in units of the LCs and site occupancies by specific atoms entering chemical formula of a considered compound. If the symmetry properties for each atomic position are known, the whole crystal lattice can be built up by repeating the unit cell in three directions with a proper application of the corresponding symmetry operations.

There are well-known methodics of experimental determination of the crystal structure from a thorough analysis of the X-ray and neutron diffraction patterns. From the theoretical point of view, it is also possible nowadays to get the structural properties of any crystal using the widely spread reliable *ab initio* methods of calculations. The rapid development and improvement of both experimental facilities and computational techniques allowed for getting trustworthy information on crystal structure of a large number of materials, which has been collected into various commercial and freely available databases. Comparing the accuracy of the experimental and theoretical methods of determination of crystal structure, it is worthwhile to note that the structural discrepancy between the theoretically calculated and experimentally deduced parameters for the same crystal typically does not exceed a few percent on average and very often is even less than 1%.

At the same time, the two above-mentioned methodics of determination of the crystal lattice structure – no matter how precise they can be in every particular case – give no opportunity to make a quick and reliable estimation and/or prediction of the structural parameters for even isostructural compounds, since all measurements and/or calculations are essentially *ad hoc* and should be repeated again for any new crystal. It is also noteworthy that both methods require sophisticated equipment and/or computational skills; in addition to that, they are expensive and time consuming.

In this connection, a simple empirical model, which encompasses a large group of isostructural materials and links together the lattice structure parameters with various characteristics of their constituting elements (e.g. ionic radii, oxidation state, electronegativity etc.) can be useful for researchers working in the experimental materials science and chemistry. The usefulness of such models originates from their simplicity and ability to make quickly reliable prediction and/or estimation of the LCs for those materials, which have not been experimentally found yet. One of possible applications of such models can be related to a choice of proper substrates – with suitable structural properties – for the thin films growth.

It is a long-time-ago established fact that the ionic radii are one of the most important parameters, responsible for the interionic separations and, as a consequence, LCs of crystals. Two other key-parameters are the electronegativity and oxidation state,[1] which can greatly affect the chemical bond properties and, finally, the interionic separation. It should be kept in mind that these parameters are the empirical quantities, which may be defined in different ways and, depending on definitions and/or degree of experimental precision; they can be given somewhat different values.

Speaking about electronegativity, we mention here that there exist several different electronegativity scales, e.g. Martynov-Batsanov scale,[2] Phillips scale,[3] Jaffe scale,[4] Allen scale[5] etc. Throughout this paper, the use is made of the Pauling electronegativity scale[6] and the Shannon's ionic radii[7] for all considered ions.

A thorough statistical analysis of a large number of crystals of a given structure can help in finding a functional relation between these or any other parameters and LCs values. The cubic crystals with their single crystal lattice parameter $a$ are a special group of materials, whose LCs can be analyzed in terms of the properties of the constituting chemical elements. Recently, several papers[8-12] dealing with the empirical modeling of the LCs for the cubic perovskite crystals were published. The linear relations between the value of $a$ and several other variables (ionic radii, number of valence electrons, and electronegativity) in various combinations were proposed and successfully tested. In a similar way, the LCs of a group of the $A_2XY_6$ cubic crystals ($A$=K, Cs, Rb, Tl; $X$=tetravalent cation, $Y$=F, Cl, Br, I)[13] and cubic pyrochlores[14] were modeled with achieving good agreement between the predicted and experimental LCs values. A recent analysis of the pyrochlore structural data[15] allowed introducing a new empirical tolerance factor for the representatives of this group of compounds. So, modeling of the crystal lattice constants and structures has never stopped and still appears to be an interesting and attractive problem fundamental and applied importance.

In the present work we consider a group of crystals with the spinel structure. This is a very large family of compounds. They are not only widely-spread in nature occurring as pure (or mixed) minerals all over the globe; the spinels are also significant in many branches of technology and science. Many spinels are typical semiconductors with a rather narrow band gap (this is true especially for spinels containing the halogen atoms as anions), whereas the oxygen-based spinels have considerably wider band gaps and thus are typical dielectrics, which can be easily doped with rare earth and transition metal ions. For example, $MgAl_2O_4$ and $ZnGa_2O_4$ doped with $Co^{2+}$ ions were shown to be promising materials for solid state lasers;[16] $Ni^{2+}$-doped $MgAl_2O_4$ was named as an active medium for the tunable infrared solid state laser.[17] The spinel-based transparent ceramics for high-energy laser systems were designed recently.[18] There are also numerous examples of doping spinel compounds with rare earth ions, e.g. $ZnAl_2O_4$:$Ce^{3+}$, $Tb^{3+}$,[19] $MgGa_2O_4$:$Pr^{3+}$,[20] $MgAl_2O_4$:$Nd^{3+}$,[21] $Dy^{3+}$, $Sm^{3+}$, $Er^{3+}$, $Eu^{3+}$, and $Tm^{3+}$ doped $MgIn_2O_4$ [22] etc. Many spinels exhibit magnetic properties, which are a subject of many research works[23-25] with practical applications in view.

The spinel-type compounds are known for a long time already, and much effort was applied to clarify and understand their structural properties.[26-31] The "classical" spinels are the ternary compounds that are described by the $AM_2X_4$ chemical formula, where $A$ and $M$ are the metals occupying the tetra- and octahedrally coordinated positions, respectively, and $X$ stands for the anion, which can be any of these elements: oxygen, sulfur, selenium, tellurium, nitrogen. There exists certain "internal degree of freedom" in distributions of the cations through the tetra- and octahedral positions; one can distinguish between the so called "normal" $A(M_2)X_4$ and "inverse" $M(AM)X_4$ distributions, where the ions in the parentheses are located at the octahedral sites.[32] Intermediate distributions can also occur, covering the whole range between the normal and inverse spinels; they can be generally described as $A_{1-\lambda}M_\lambda(A_\lambda M_{2-\lambda})X_4$ with $\lambda$ representing the degree of inversion ($\lambda$=0 for the normal spinels and $\lambda$=1 for the inverse ones). The anion fractional coordinate $u$ in the spinel structure was shown to depend strongly on the cation inversion parameter.[33]

It has been demonstrated that the octahedral and tetrahedral bond lengths (i.e. the interionic distances in the *A*-*X* and *M*-*X* pairs, respectively) in the spinel structure can be used to predict the lattice constant *a* and the anion positional coordinate *u*.[27] Several works also have been published that stress out existing correlations between various physical properties of spinels and ionic radii of the constituting ions. Thus, a relation between the magnetic and ionic properties of spinels with the ionic radii of cations and anions was discussed earlier.[34] Systematics of some spinel compounds based on the ionic radii of the constituting ions and geometrical factors of the spinel's crystal lattice structure was suggested in Refs. [27, 35, 36] In particular, a comprehensive data base of the spinel-type compounds was collected in Ref. [36].

In the present work we propose a new semi-empirical approach, which allowed us to model and describe the lattice parameters of ternary and binary spinels. The model treats the ionic radii and electronegativities of the constituting ions forming the spinel crystal lattice as the main factors to determine the value of the lattice parameter. Inclusion of electronegativities into our model extends and refines previous attempts of modeling spinel crystal lattices,[26-31, 35, 36] which were based on the geometrical factors only, such as ionic radii and interionic separations.

The reason for addition of electronegativity is due to the fact that the purely geometric consideration based on the ionic radii alone cannot explain why some compounds, although built up from the ions with equal ionic radii, have, nevertheless, different LCs. One example of this kind is the pair of the $Cs_2GeF_6$ and $Cs_2MnF_6$ crystals: although the ionic radii of $Ge^{4+}$ and $Mn^{4+}$ (the only different ions in these compounds) are equal, their LCs are slightly different.[13] Inclusion of the electronegativity as one of those parameters, which determine the bonding properties, can help in handling this issue and refine further those models, based entirely on the geometrical considerations and ionic radii, when the ions in a crystal lattice are treated as incompressible hard spheres.

The model developed and described in the present paper was tested by considering a group consisting of 185 binary and ternary stoichiometric $AM_2X_4$ spinel compounds, which can be divided into four sub-groups depending on the anion *X*. These sub-groups are conditionally referred to in the present paper as the oxides (*X*=O, 83 compounds), sulfides (*X*=S, 56 compounds), selenides/tellurides (25 selenides and 3 tellurides, *X*=Se,

Te, 28 compounds in total), and nitrides (*X*=N, 18 compounds). All the nitride spinels included into the present model were reported only theoretically, using the *ab initio* calculation techniques for optimizing their crystal structure, and, as such, they stand apart from other considered compounds.

The main aim of the performed analysis was to find simple empirical rules for a proper description of lattice parameters of the experimentally found spinels and predict the LCs of those new materials, experimentally not found yet, which can be, in principle, synthesized. Among the experimental spinel-type compounds are those synthesized at high pressure (metastable), simple, inverse and binary spinels; putting them together gives only a general view of the structural properties of spinels.

The linear relation between the LCs, ionic radii and electronegativities of the constituting ions allowed us to calculate the LCs of already existing spinels with an average deviation between the experimental data and our model estimations of less than 1 %: the fact, which serves as a firm justification of the validity, applicability and potential predictive abilities of the derived equation. A closer look at certain relations, which link together the ionic radii and electronegativity of existing stable spinels, helped us to reveal certain limits (or typical ranges) for variations of these parameters, which may set up the boundaries of stability of spinel compounds. This should be of paramount importance in a search for new not-synthesized yet materials, since such conditions, once established, effectively reduce the number of possible combinations of chemical elements to be considered potentially suitable.

## 2. Results and discussion

All chosen compounds crystallize in the Fd-3m space group (No. 227), with eight formula units in one unit cell. The unit cell of $MgAl_2O_4$ – a classical representative of the spinel group – is shown in Fig. 1.

In this material the oxygen ions form a cubic close packing; the Mg ions occupy 1/8 of the available 4-fold coordinated tetrahedral positions, whereas the Al ions – ½ of the available 6-fold coordinated octahedral sites.[37]

Table 1 collects the LCs values found in the literature. The vast majority of the data included into Table 1 correspond to the experimental structural studies of the synthesized spinel compounds. Some structural data were taken from the theoretical works on *ab initio* studies of the spinel compounds (followed by an asterisk in Table 1), and no corresponding experimental data were found. The set of the data collected in Table 1 is considerably extended if compared to that one from Ref. [36]; more recent literature data were used when compiling Table 1.

The ionic radii of all ions were taken from Ref. [7] and the Pauling electronegativities were those listed in Ref.[38]. The LCs were taken mainly from the Inorganic Crystal Structure Database (ICSD) [39] and from some additional publications, explicitly cited in the Table. All compounds in Table 1 are sorted as follows: oxides, sulfides, selenides, tellurides, nitrides. In each of these groups the alphabetical ordering was used to list all entries.

As it can be found from this table, the oxide spinels have the LCs in the range between 8.044 Ǻ ($SiNi_2O_4$) to 9.26 Ǻ ($MoAg_2O_4$); those of the sulfide spinels vary from 9.4055 Ǻ ($Co_3S_4$) to 11.26 Ǻ ($CdDy_2S_4$), and those of the selenide spinels are in the interval from 10.20 Ǻ ($Co_3Se_4$) to 11.647 Ǻ ($CdDy_2Se_4$). Three telluride spinels, whose structural data were found in the ICSD, are $AgCr_2Te_4$, $CdDy_2Te_4$ and $CuCr_2Te_4$ with the LCs of 11.371, 11.38 and 11.26 Ǻ, correspondingly. A group of the nitride spinels has the LCs in the range from 7.2867 Ǻ (c-$SiC_2N_4$) to 9.1217 Ǻ (c-$Zr_3N_4$). So, the *total* range for the LCs values presented in the table covers a wide interval from 7.2867 Ǻ to 11.647 Ǻ - more than 3.5 Ǻ. We also emphasize again that the most of the nitride spinels listed in Table 1 were obtained theoretically only, using the ab initio calculations. The binary spinels, such as $Co_3O_4$ and $Fe_3O_4$, are listed as $CoCo_2O_4$ and $FeFe_2O_4$, to distinguish between the doubly and triply positively charged ions at the tetra- and octahedral sites, respectively.

In a vast majority of the selected spinels (except for nitrides) the oxidation state of the ion located at the tetrahedral site is “+2”, the oxidation state of the octahedrally coordinated ion is “+3”, and the oxidation state of the anion is “-2”. These oxidation states appear as a consequence of the partial occupancy of the tetra- and octahedral sites and are characteristic of normal spinels. The exceptions are as follows: i) *A*=Ge, Si, Sn (oxidation state +4, oxidation state of the *M* cation +2); ii) *A*=Li (oxidation state +1, oxidation state of the *M* cation +3.5, obtained as a one-to-one mixture of the cations in the oxidation states +3 and +4; iii) *A*=Mo, W (oxidation state +4 or +6, then the *M* cation has the oxidation state +2 or +1, respectively).

At first, all the LCs from Table 1 were fitted to the linear function of the following variables: two sums of ionic radii ($R_A + R_X$), ($R_M + R_X$) and two differences of electronegativities $(\chi_X - \chi_M), (\chi_X - \chi_A)$. The choice of these variables seems to be quite natural, since both *A* and *M* ions are surrounded by the *X* ions. The sum of ionic radii of two neighboring ions can be taken as an interionic separation. This is, of course, an approximation only, since it is based on a model representing both atoms as rigid incompressible spheres. The difference of electronegativities of two neighboring ions is a characteristic of degree of ionicity (covalency) of the chemical bond: the greater is such difference, the more ionic the bond is. For pure covalent bonds, like in the diatomic molecules of hydrogen or oxygen, the difference of electronegativities of the atoms forming the chemical bond is obviously zero; in the case of heteropolar bonds such difference is not zero, which indicates certain ionicity of such bonds.

The performed least square fit resulted in the following linear function, describing the LCs of the chosen crystals:

$$a_{calc} = 1.20740(R_A + R_X) + 2.67682(R_M + R_X) + 0.11573(\chi_X - \chi_M) + 0.10840(\chi_X - \chi_A) + 0.26705 \quad (1)$$

In this equation the ionic radii $R_A, R_M, R_X$ and the calculated LCs $a_{calc}$ are expressed in Å; the electronegativities $\chi_A, \chi_M, \chi_X$ are dimensionless. Therefore, the fitting coefficients before the ionic radii sums ($R_A + R_X$) and ($R_M + R_X$) are dimensionless, whereas the coefficients before the electronegativities differences $(\chi_X - \chi_M), (\chi_X - \chi_A)$ have the dimension of Å.

The correlation between the LCs taken from the literature and calculated by Eq. (1) is shown in Fig. 2. In addition, the numerical results obtained from Eq. (1) are also given in Table 1, along with the absolute (in Å) and relative errors (in%) in comparison with the experimental data. The straight line in Fig. 2 has a slope equal to unity and corresponds to the perfect one-to-one match between the experimental and predicted LCs. Since the nitride-based spinels were predicted theoretically using the ab initio calculations, they all are shown by empty symbols, to separate them clearly from the experimentally reported spinels.

Despite simplicity of Eq. (1), it already gives a reasonable estimate of the lattice parameter for the most of the considered spinels. The average error is 1.45 %; the root-mean-square deviation between the calculated and experimental LCs is 0.182 Å. The greatest error between the experimental and predicted LCs is 8.79 % for $CdDy_2Te_4$, which also may give some hint of the necessity to re-consider the corresponding experimental LC of this compound.

Among those 185 spinels, considered in the presented model, for 84 compounds the relative error does not exceed 1.0 %; for 62 crystals the relative error is in the range from 1.0 % to 2.0 %; for 22 crystals the relative difference between the calculated and experimental values is from 2.0 % to 3.0 %, for 9 of them the relative error varies from 3.0 % to 4.0 %, for 2 of them the relative error is from 4 % to 5 %, for 5 - from 5 % to 6 %, and for 1 crystal it is equal to 8.79 %.

However, the fact that there was found such a big error (8.79 %) between our model and literature data for the $CdDy_2Te_4$ spinel induced us to refine the model and treat separately various groups of spinels, depending on the anion, like oxides, sulfides, selenides together with tellurides, and nitrides, because in Eq. (1) we do not distinguish between these groups of spinels with different anions.

Then new linear fits of the LCs for oxides, sulfides, selenides/tellurides and nitrides were obtained as follows:

$$a = 1.27084(R_A + R_X) + 2.49867(R_M + R_X) + 0.08640(\chi_X - \chi_M) + 0.05141(\chi_X - \chi_A) + 0.60340 \quad \text{(oxides)} \quad (2)$$

$$a = 1.51899(R_A + R_X) + 2.90926(R_M + R_X) + 0.34215(\chi_X - \chi_M) + 0.40573(\chi_X - \chi_A) - 1.55548 \quad \text{(sulfides)} \quad (3)$$

$$a = 1.17546(R_A + R_X) + 2.01022(R_M + R_X) + 0.35765(\chi_X - \chi_M) + 0.44993(\chi_X - \chi_A) + 1.66629 \quad \text{(selenides/tellurides)} \quad (4)$$

$$a = 1.72112(R_A + R_X) + 2.22417(R_M + R_X) - 0.00447(\chi_X - \chi_M) + 0.17300(\chi_X - \chi_A) + 0.47411 \quad \text{(nitrides)} \quad (5)$$

Fig. 3 shows the results of applications of Eqs. (2) - (5) to the considered groups of spinels. With these new equations, agreement between the predicted and experimental/ab initio (the latter is related to the nitride spinels) data on the LCs has been improved considerably (compare with Fig. 2). The averaged deviation between the calculated and literature LCs is now 0.90 %. With those individual fittings for each group of spinels, LCs of 122 compounds are described by the absolute error less than 1 %; for 43 the absolute error is between 1 % and 2 %; for 12 – between 2 % and 3 %; for 7 – between 3 % and 4 %, and for 1 – 5.06 % (the same $CdDy_2Te_4$). The root-mean-square deviation between the calculated and experimental LCs is now 0.10 Å for oxide spinels, 0.145 Å for sulfide spinels, 0.187 Å for selenide/telluride spinels, and 0.069 Å for nitride spinels.

It can be noticed immediately that the nitride spinels represent a somewhat special class of compounds, since for them the coefficient at $\left(R_A + R_X\right)$ is considerably greater and the coefficient at $\left(R_M + R_X\right)$ is considerably smaller than for the remaining spinel groups. It is also easy to see that the role played by the electronegativities difference is not the same in these groups: the coefficients at $\left(\chi_X - \chi_M\right)$ and $\left(\chi_X - \chi_A\right)$ are very small for oxide spinels, whereas their values are much greater in the cases of the sulfide and selenide/telluride spinels.

One of possible factors, which is extremely hard – if possible at all – to model, is that many spinels are described as the structures, which are intermediate between the normal and inverse spinels, with quite different occupations numbers of the tetra- and octahedral positions. As a rule, the majority of the tetrahedral sites are occupied by the *A* ions, and the majority of the octahedral sites – by the *M* ions. For example, in $CuAl_2O_4$ the tetrahedral sites are occupied as follows: 64 % - by the $Cu^{2+}$ ions and 36 % by the $Al^{3+}$ ions, whereas 82 % of the octahedral sites are taken by the $Al^{3+}$, and 18 % - by the $Cu^{2+}$ ions. In $ZnAl_2O_4$ 98.4 % of the tetrahedral sites are occupied by the $Zn^{2+}$ ions and the remaining 1.6 % by the $Al^{3+}$ ions. At the same time, in this spinel 99.2 % of the octahedral sites are taken by the $Al^{3+}$, and only 0.8 % - by the $Zn^{2+}$ ions. An almost opposite example is $CuCo_2O_4$: the tetrahedral sites are occupied by both $Cu^{2+}$ and $Co^{3+}$ ions with the 1:1 ratio, whereas the 25 % of the octahedral sites are occupied by the $Cu^{2+}$ ions and 75 % - by the $Co^{3+}$ ions.

These examples show a rather random character of variation of the tetra-/octahedral sites' occupation ratio. Therefore, in our model we assumed that the *A* ions are *always* at the tetrahedral sites (except for the $RhM_2S_4$ compounds), whereas the *M* ions are *always* at the octahedral ones (the normal spinel structure). However, even with this assumption the developed model gives an adequate description of the distribution of the LCs values in the spinels' group.

One additional reason, which without any doubts contributes to the discrepancy between the estimated LCs from our model and those from the literature, is associated with the different experimental conditions at which the data are taken. Quite often, for the same compound a search can reveal several experimental LCs values, which may differ by several percent. Obviously, the experimental conditions (temperature, pressure) and crystal growth procedure (which may or may not lead to contamination of the samples by some unwanted impurities) are those factors, which, on one hand, to a large extent determine the degree of precision of the reported experimental LCs, but on the other hand, they are extremely difficult to be evaluated in order to choose the most reliable experimental result. Having realized this, we tried to select the experimental structural data obtained at ambient pressure and at room (or low, where available) temperature.

Successful modeling of the lattice parameters of the existing compounds allows to check the predictive power of the model. Table 2 below collects the structural data for three groups of spinels: $ZnM_2O_4$, $CdM_2S_4$, and $CdM_2Se_4$, with *M*=Sc, Y, La-Lu. Only very few experimental data on some members of the chosen group do exist and have been reported so far; however, the most part of these compounds have not been synthesized yet. The predicted lattice constants for these potentially new spinels are given in the table; they were obtained by using Eqs. (3) – (5). We note here that for the selenides spinels we have used a slightly modified equation, which was obtained by excluding the tellurides compounds (since there are only three of them) from the fit:

$$a = 1.71560(R_A + R_X) + 2.25828(R_M + R_X) + 0.25786(\chi_X - \chi_M) + 0.80466(\chi_X - \chi_A) - 0.67582$$

The calculated LCs values from Table 2 can be checked if the spinels mentioned there would be synthesized experimentally.

Fig. 4 allows for visualizing a linear trend, which exists between the predicted LCs in Table 2 and ionic radii of the *M* ions (*M*=Sc, Y, La-Lu). The "lanthanide contraction" (a decrease of the trivalent lanthanide ionic radii when going from La to Lu) is accompanied by a decrease of the LCs. The lines shown in Fig. 4 are the guides to the eye only; we refrained from performing a linear fit of these data points (which might be done, of course), since it would eliminate an influence of anions (O, S, Se) and electronegativities on the calculated result.

Nevertheless, a linear variation of the LCs in each of the considered groups, which agrees with the Vegard's law, can serve as an additional argument favoring the estimations of LCs for those not reported yet rare earth-based spinels.

## 3. Stability ranges of ternary spinels

Careful consideration of the properties of the constituting ions in ternary spinels can help in establishing limits for the stable/unstable compounds, thus effectively narrowing down the search space for the new materials. Although various combinations of the characteristics of crystal lattice ions can be constructed, one of those, which eventually turned out to be most useful, is the bond stretching force constant[35]

$$K_{AM} = \frac{\chi_A \chi_M}{(R_A + R_X)^2 + (R_M + R_X)^2 + 1.155(R_A + R_X)(R_M + R_X)} , \qquad (6)$$

where all quantities have been defined above. This quantity, as emphasized by Kugimiya and Steinfink,[35] was extremely efficient for indicating the stability ranges for various $AB_2O_4$ structures, including the spinel and olivine phases. Fig. 5 shows dependence of the experimental lattice constant of all spinels from Table 1 on the $K_{AM}$ value. It can be immediately seen from the figure that the group of spinels differing by the anions occupy different regions in that diagram. The oxides, for example, are well separated from other compounds. The nitride spinels are scattered over a wide area, but this can be explained by instability of the nitride spinels and by the facts that many of those nitride compounds were reported theoretically only.

Fig. 6 shows the scattered plot of the experimental lattice parameter versus a non-dimensional ratio of the sums of ionic radii $(R_A + R_X)/(R_M + R_X)$. This diagram imposes certain limits for this ratio: thus, if the upper limit is about 1.2 for all compounds, the lower limit of $(R_A + R_X)/(R_M + R_X)$ is about 0.88 for sulfides/selenides/tellurides, and about 0.5 for oxides. So, the generalization of this diagram can be put forward as follows: if the atomic radii in the $AM_2X_4$ ternary spinels are concerned, the $(R_A + R_X)/(R_M + R_X)$ ratio is expected to be between 0.5 and 1.2, and existence of a stable ternary spinels with ionic radii not satisfying these conditions, seems to be unlikely, at least, at the ambient conditions.

As an intrinsic check for the reliability of our predicted lattice constants of the rare-earth-based oxide, sulfide, and selenide spinels from Table 2, we included the corresponding data points (shown by the empty symbols to make them easily distinguishable from the rest of the figure) into Fig. 6. These predicted compounds are all in the above-suggested stability range, since the above-introduced $(R_A + R_X)/(R_M + R_X)$ ratio for all of them is between 0.8-0.95 (oxides) and 0.9-1.0 (sulfides, selenides).

We also present in Fig. 7 another scatter plot, which suggests a certain correlation between the sum of electronegativities $\chi_A + \chi_M + \chi_X$ and the sum of ionic radii $R_A + R_M + R_X$ in the group of 185 considered spinels. An important observation to be made is that the value of $R_A + R_M + R_X$ about 3 Å is a border between the oxide spinels with $R_A + R_M + R_X$ < 3 Å and sulfides, selenides, tellurides with $R_A + R_M + R_X$ >3 Å. One oxide spinel $MoAg_2O_4$ and one telluride spinel $CdDy_2Te_4$ clearly fall out from the corresponding groups, and this can be a hint to certain experimental inaccuracies in determination of their LCs or certain questions regarding their stability. Another possible reason for that can be related to a large difference between electrical charges of Mo and Ag ions (+6 and +1, respectively), whereas in other considered spinels the cations' charges are +2 and +3. As far as $CdDy_2Te_4$ is concerned, it should be mentioned that the experimental data on this compound are very scarce, and can hardly be verified. The region of the stable oxide spinels is characterized by the sum of electronegativities in the range from 6 to 7.7 and an averaged sum of three ionic radii $R_A + R_M + R_X$ about 2.6 Å. The sulfide and selenide spinels cannot be clearly separated in this diagram; their representing data points occupy the area with the electronegativity sum between 5 and 6.8 and the ionic radii sum between 3 and 3.7 Å, with the averaged value of about 3.3 Å. The possibility to group the representing data points of different spinels in Fig. 7 into various regions of stability can help in choosing suitable chemical elements for new spinels.

Finally, Fig. 8 presents a well-determined correlation between the unit cell volumes of the considered spinel compounds and the sums of volumes of individual ions (the latter are considered as the hard spheres with the Shannon ionic radii). The relation between these quantities is a linear one, as shown by the linear fits with explicitly given equations of those fits. As follows from Fig. 8, there are certain lower and upper limits, within which the spinels of certain types (oxides, sulfides, selenides, tellurides, nitrides) can exist.

For example, the experimental volume of the unit cell of the oxide spinels varies between 500 and 800 Å$^3$ with a sum of volumes of individual ions in a unit cell less or about 400 Å$^3$. The experimental volume of the unit cell of the sulfur spinels is in the range ~800 and 1500 Å$^3$ due to a greater ionic radius of sulfur if compared with that of oxygen; a sum of volumes of individual ions in a unit cell of sulfide spinels is less or about 900 Å$^3$. The sum of individual volume of ions in one unit cell of the selenide spinels is about 1100 Å$^3$, whereas the experimental volumes of one unit cell are confined within the 1000 – 1600 Å$^3$ range.

An observation can be made that the ratio of the experimental volume of one unit cell to the sum of volumes of ions in such a cell is decreasing when going from oxide spinels to sulfides and further to selenides. In other words, in more covalent spinels, such as sulfides and selenides, the ions are packed more closely, and the fraction of the empty space between the ions is decreasing. The nitride spinels in this sense are more ionic and share more resemblance with the oxygen-based spinels. However, the circumstance that many of the nitrides mentioned in the present paper were obtained only theoretically prevents us from making any further conclusions regarding their stability.

The group of the telluride spinels, which consists only of three members, is also included into Fig. 8 for the sake of completeness of the undertaken study. Tellurium is the largest anion among all considered in the present paper, and the sum of the individual ions volumes in a unit cell of the tellurium-based spinels is practically equal to the experimental volume of a unit cell.

The dashed lines in Fig. 8 are the lower and upper boundaries, within which all the studied compounds are located; these limiting lines determine the filling factors (ratio of the sum of volumes of the constituting ions to that of the unit cell). For the oxide spinels such filling factor varies from 0.52 to 0.72 with the average value of 0.64. For the group of the sulfide and selenide compounds this range is shifted towards greater values: 0.63 – 1.03 with the average value of 0.80 and 0.69 – 1.00 with the average value of 0.79 for the sulfide and selenide spinels, respectively. The filling factor is about 1 for three tellurium-based spinels. As a guide to the eyes, we also plotted in Fig. 8 a straight line with a slope equal to 1, which would mean that the experimental volume of a unit cell is equal to the sum of volumes of individual ions – such a condition is practically never met.

As for the filling factor, in the system of equally sized spheres the dense packing corresponds to filling of 0.81. In the system of spheres of two or more different sizes, dense packing may mean a filling factor closer to 0.9. The value of about 1 (seen in Fig. 8) means just that the bond lengths in the crystal are shorter than those resulting from a simple hard spheres model. If we would assume that the shortening results in volume reduction of 10%, this means that the bonds are shorter by about 3% for the spinel compounds located in Fig. 8 at the line corresponding to the filing factor 1.

It can be anticipated that the spinel compounds (including those, which are not synthesized yet), whose representing points would appear in Fig. 8 outside of the region bordered by the two dashed straight lines, would be unstable or would require special conditions for synthesis (high pressure, for example).

## 4. Conclusions

We propose in the present paper a simple model, which allows for establishing a simple correlation between the lattice constant, ionic radii and electronegativities of the constituting ions in the case of the ternary spinel compounds $AM_2X_4$, where $A$ and $M$ are the metals occupying the tetra- and octahedrally coordinated positions, respectively, and $X$ stands for the anion. A linear equation was obtained that links together the lattice constant with sums of the pairs of ionic radii ($R_A + R_X$), ($R_M + R_X$) and differences of pairs of electronegativities $\left(\chi_X - \chi_M\right)$, $\left(\chi_X - \chi_A\right)$. The developed model has been tested in a group of 185 spinels, whose structural data were found in the literature. The fitting was performed separately for the spinels with different anions (oxygen, sulfur, selenium/tellurium, nitrogen). The model's equation yields good agreement between the experimental and predicted lattice constants, with an average error of 0.90 % only, for 122 spinels out of 185 considered compounds the relative error between the experimental and calculated lattice constants is less than 1 %. The model proposed in the present paper is an empirical one, and the choice of its main parameters – ionic radii and electronegativities – looks to be a natural choice, since these factors to a large extent and in the first approximation determine the interionic separations, size of the interstitial positions in the crystal lattice and, finally, the lattice constants themselves. It should be also emphasized that the coefficients in Eqs. (1)-(5), obtained from the linear fit to the experimental data, depend on the scale of electronegativities and ionic radii, as has been mentioned clearly in the introduction. Our results held true for the Pauling electronegativities and Shannon radii.

A close look at the collected in the present work experimental and modeled lattice constants reveal that the chemical and physical properties of the constituting chemical elements can also significantly contribute to the deviation between the model and experiment. Thus, the spinels with transition metal ions, such as Mn, Fe, Co, Ni, which exhibit magnetic properties due to the presence of the unfilled 3d electron shell and its active participation in chemical bonding, are those compounds whose modeled lattice constants in many cases deviate more significantly from the experimental results. This circumstance may be a hint for a further development of the present model, which can be a future perspective.

Careful consideration of the interplay between the experimental lattice constants and/or ionic radii, bond stretching force constant, sum of volumes of the constituting ions, their ionic radii and electronegativities allows us to identify the certain regions of stability, within which the stable spinel compounds can be expected to exist. The obtained trends were represented by the two-dimensional plots; their meaning was discussed in the text. The main application of those plots, as it is deemed now, would be to narrow down the search for new spinels by choosing those potential compounds whose representative points would fall down within the domains of existing stable compounds.

We believe that the obtained empirical dependence of the lattice constant on the ionic radii and electronegativity difference, expressed by Eqs. (1) - (5) from this paper, will be helpful for the chemists and materials scientists, since it gives an opportunity to estimate in a very simple and efficient way the lattice constants for new ternary compounds with the spinel structure. It is essential that the developed here model not only takes into account the ionic radii as the main geometrical factors to determine the lattice constant, but accounts – at least, partially – for a difference in chemical properties of the constituting ions by considering explicitly the difference of electronegativities of nearest neighbors making chemical bonds. We also hope that the results obtained in the present paper can be useful for meaningful guided choice of chemical elements for a synthesis of new spinel compounds.

**Acknowledgements**

This study was supported the bilateral project between the Estonian and Polish Academies of Sciences in the years 2010-2012 and 2013-2015 and the project number DEC-2012/07/B/ST5/02080 of National Science Center of Poland. M.G.B. appreciates financial support from the European Union through the European Regional Development Fund (Centre of Excellence "Mesosystems: Theory and Applications", TK114), European Internationalization Programme DoRa, Marie Curie Initial Training Network LUMINET, grant agreement No. 316906 and the Programme for the Foreign Experts offered by Chongqing University of Posts and Telecommunications. The authors thank Prof. W. Paszkowicz (Institute of Physics, Polish Academy of Sciences) for fruitful discussions and critical remarks.

Table 1. Experimental and predicted (this work) lattice constants of various spinel compounds $AM_2X_4$. Compounds whose lattice constants were *ab initio* calculated earlier are marked with an asterisk.

| No. | ICSD No. or Ref. | Composition | LC exp., Å | Eq. (1) LC calc.,Å | Eq. (1) Abs. error (exp-calc), Å | Eq. (1) Relative error,% | Eqs. (2) - (5) LC calc.,Å | Eqs. (2) - (5) Abs. error (exp-calc), Å | Eqs. (2) - (5) Relative error,% |
|---|---|---|---|---|---|---|---|---|---|
| 1 | [40] | $CdAl_2O_4$ | 8.355 | 8.40265 | -0.04765 | 0.57027 | 8.38145 | -0.02645 | 0.31658 |
| 2 | 37428 | $CdCr_2O_4$ | 8.567 | 8.61101 | -0.04401 | 0.51366 | 8.57702 | -0.01002 | 0.11696 |
| 3 | 66133 | $CdFe_2O_4$ | 8.7089 | 8.67164 | 0.03726 | 0.4279 | 8.6373 | 0.0716 | 0.82215 |
| 4 | 43743 | $CdGa_2O_4$ | 8.59 | 8.60703 | -0.01703 | 0.19824 | 8.57656 | 0.01344 | 0.15646 |
| 5 | 4118 | $CdIn_2O_4$ | 9.166 | 9.09233 | 0.07367 | 0.80374 | 9.02891 | 0.13709 | 1.49564 |
| 6 | 28954 | $CdRh_2O_4$ | 8.73 | 8.67309 | 0.05691 | 0.65188 | 8.64839 | 0.08161 | 0.93482 |
| 7 | 28961 | $CdV_2O_4$ | 8.695 | 8.6814 | 0.0136 | 0.15645 | 8.64208 | 0.05292 | 0.60863 |
| 8 | 77743 | $CoAl_2O_4$ | 8.0968 | 8.14057 | -0.04377 | 0.54057 | 8.11751 | -0.02071 | 0.25578 |
| 9 | [36] | $CoCo_2O_4$ | 8.0835 | 8.31008 | -0.22658 | 2.80303 | 8.28159 | -0.19809 | 2.45055 |
| 10 | 69503 | $CoCr_2O_4$ | 8.333 | 8.34893 | -0.01593 | 0.19114 | 8.31309 | 0.01991 | 0.23893 |
| 11 | [36] | $CoFe_2O_4$ | 8.35 | 8.40956 | -0.05956 | 0.71327 | 8.37336 | -0.02336 | 0.27976 |
| 12 | 77744 | $CoGa_2O_4$ | 8.3229 | 8.34495 | -0.02205 | 0.26496 | 8.31262 | 0.01028 | 0.12351 |
| 13 | 109301 | $CoRh_2O_4$ | 8.495 | 8.41101 | 0.08399 | 0.98865 | 8.38445 | 0.11055 | 1.30135 |
| 14 | [36] | $CoV_2O_4$ | 8.4070 | 8.41932 | -0.01232 | 0.14656 | 8.37815 | 0.02885 | 0.34317 |
| 15 | 172130 | $CuAl_2O_4$ | 8.0778 | 8.12633 | -0.04853 | 0.60075 | 8.10378 | -0.02598 | 0.32162 |
| 16 | [36] | $CuCo_2O_4$ | 8.054 | 8.29584 | -0.24184 | 3.00274 | 8.26785 | -0.21385 | 2.6552 |
| 17 | [36] | $CuCr_2O_4$ | 8.2700 | 8.33469 | -0.06469 | 0.78218 | 8.29935 | -0.02935 | 0.3549 |
| 18 | [36] | $CuFe_2O_4$ | 8.369 | 8.39532 | -0.02632 | 0.31445 | 8.35962 | 0.00938 | 0.11208 |
| 19 | 61028 | $CuGa_2O_4$ | 8.298 | 8.33071 | -0.03271 | 0.39419 | 8.29889 | -8.9E-4 | 0.01073 |
| 20 | 27922 | $CuMn_2O_4$ | 8.33 | 8.42772 | -0.09772 | 1.17313 | 8.38382 | -0.05382 | 0.6461 |
| 21 | [36] | $CuRh_2O_4$ | 8.29 | 8.39677 | -0.10677 | 1.28796 | 8.37072 | -0.08072 | 0.9737 |
| 22 | [36] | $FeAl_2O_4$ | 8.149 | 8.20636 | -0.05736 | 0.70388 | 8.18363 | -0.03463 | 0.42496 |
| 23 | 98551 | $FeCo_2O_4$ | 8.242 | 8.37587 | -0.13387 | 1.62428 | 8.3477 | -0.1057 | 1.28246 |
| 24 | 43269 | $FeCr_2O_4$ | 8.378 | 8.41472 | -0.03672 | 0.43827 | 8.3792 | -0.0012 | 0.01432 |
| 25 | [36] | $FeFe_2O_4$ | 8.394 | 8.47535 | -0.08135 | 0.96912 | 8.43947 | -0.04547 | 0.5417 |
| 26 | 28285 | $FeGa_2O_4$ | 8.363 | 8.41074 | -0.04774 | 0.57087 | 8.37873 | -0.01573 | 0.18809 |
| 27 | [36] | $FeTi_2O_4$ | 8.500 | 8.57583 | -0.07583 | 0.89214 | 8.52699 | -0.02699 | 0.31753 |
| 28 | 28666 | $FeMn_2O_4$ | 8.51 | 8.50775 | 0.00225 | 0.02639 | 8.46366 | 0.04634 | 0.54454 |
| 29 | 109150 | $FeNi_2O_4$ | 8.288 | 8.34563 | -0.05763 | 0.69538 | 8.32012 | -0.03212 | 0.38755 |
| 30 | 28962 | $FeV_2O_4$ | 8.543 | 8.48511 | 0.05789 | 0.67762 | 8.44426 | 0.09874 | 1.1558 |
| 31 | 69497 | $GeCo_2O_4$ | 8.318 | 8.42796 | -0.10996 | 1.32189 | 8.37077 | -0.05277 | 0.63441 |
| 32 | [36] | $GeFe_2O_4$ | 8.411 | 8.52743 | -0.11643 | 1.38427 | 8.46254 | -0.05154 | 0.61277 |
| 33 | 1086 | $GeMg_2O_4$ | 8.2496 | 8.427 | -0.1774 | 2.15044 | 8.35755 | -0.10795 | 1.30855 |
| 34 | [36] | $GeNi_2O_4$ | 8.2210 | 8.27726 | -0.05626 | 0.68432 | 8.23075 | -0.00975 | 0.1186 |
| 35 | [41] | $HgCr_2O_4$ | 8.658 | 8.79473 | -0.13673 | 1.57926 | 8.78984 | -0.13184 | 1.52275 |
| 36 | [36] | $LiMn_2O_4$ | 8.2460 | 8.39635 | -0.15034 | 1.82325 | 8.31161 | -0.06561 | 0.79566 |
| 37 | [36] | $LiV_2O_4$ | 8.22 | 8.44865 | -0.22865 | 2.78167 | 8.36217 | -0.14217 | 1.72956 |
| 38 | [36] | $MgAl_2O_4$ | 8.0832 | 8.19028 | -0.10708 | 1.32478 | 8.13411 | -0.05091 | 0.62982 |
| 39 | [36] | $MgCo_2O_4$ | 8.1070 | 8.3598 | -0.2528 | 3.11828 | 8.29818 | -0.19118 | 2.35821 |
| 40 | 171106 | $MgCr_2O_4$ | 8.3329 | 8.39864 | -0.06574 | 0.78897 | 8.32968 | 0.00322 | 0.03864 |
| 41 | 172279 | $MgFe_2O_4$ | 8.36 | 8.45927 | -0.09927 | 1.18749 | 8.38996 | -0.02996 | 0.35837 |
| 42 | 37359 | $MgGa_2O_4$ | 8.280 | 8.39467 | -0.11467 | 1.38488 | 8.32922 | -0.04922 | 0.59444 |
| 43 | 24231 | $MgIn_2O_4$ | 8.81 | 8.87997 | -0.06997 | 0.79419 | 8.78157 | 0.02843 | 0.3227 |
| 44 | 109299 | $MgRh_2O_4$ | 8.53 | 8.46073 | 0.06927 | 0.81208 | 8.40105 | 0.12895 | 1.51172 |
| 45 | 28324 | $MgTi_2O_4$ | 8.474 | 8.55976 | -0.08576 | 1.01201 | 8.47748 | -0.00348 | 0.04107 |
| 46 | 60412 | $MgV_2O_4$ | 8.42 | 8.46904 | -0.04904 | 0.58239 | 8.39474 | 0.02526 | 0.3 |
| 47 | 157282 | $MnAl_2O_4$ | 8.2104 | 8.27293 | -0.06253 | 0.76164 | 8.23615 | -0.02575 | 0.31363 |
| 48 | 31161 | $MnCr_2O_4$ | 8.437 | 8.48129 | -0.04429 | 0.52499 | 8.43172 | 0.00528 | 0.06258 |
| 49 | 28517 | $MnFe_2O_4$ | 8.511 | 8.54192 | -0.03092 | 0.36333 | 8.49199 | 0.01901 | 0.22336 |
| 50 | 17067 | $MnGa_2O_4$ | 8.4577 | 8.47732 | -0.01962 | 0.23194 | 8.43125 | 0.02645 | 0.31273 |
| 51 | 24999 | $MnIn_2O4$ | 9.007 | 8.96262 | 0.04438 | 0.49276 | 8.88361 | 0.12339 | 1.36993 |
| 52 | 109300 | $MnRh_2O_4$ | 8.613 | 8.54338 | 0.06962 | 0.80832 | 8.50309 | 0.10991 | 1.27609 |
| 53 | 22383 | $MnTi_2O_4$ | 8.6 | 8.64241 | -0.04241 | 0.4931 | 8.57952 | 0.02048 | 0.23814 |
| 54 | 109148 | $MnV_2O_4$ | 8.52 | 8.55169 | -0.03169 | 0.3719 | 8.49678 | 0.02322 | 0.27254 |
| 55 | 36187 | $MoAg_2O_4$ | 9.26 | 9.44357 | -0.18357 | 1.98241 | 9.34341 | -0.08341 | 0.90076 |
| 56 | 21114 | $MoFe_2O_4$ | 8.509 | 8.42357 | 0.08543 | 1.00398 | 8.41044 | 0.09856 | 1.1583 |
| 57 | 44523 | $MoNa_2O_4$ | 9.108 | 9.28192 | -0.17392 | 1.90949 | 9.15768 | -0.04968 | 0.54545 |
| 58 | 21117 | $NiAl_2O_4$ | 8.045 | 8.1011 | -0.0561 | 0.69727 | 8.07785 | -0.03285 | 0.40833 |
| 59 | 24211 | $NiCo_2O_4$ | 8.114 | 8.27061 | -0.15661 | 1.93011 | 8.24192 | -0.12792 | 1.57653 |
| 60 | 84376 | $NiCr_2O_4$ | 8.3155 | 8.30945 | 0.00605 | 0.07271 | 8.27342 | 0.04208 | 0.50604 |
| 61 | [36] | $NiFe_2O_4$ | 8.3250 | 8.37008 | -0.04508 | 0.54155 | 8.33369 | -0.00869 | 0.10438 |
| 62 | 27903 | $NiGa_2O_4$ | 8.258 | 8.30548 | -0.04748 | 0.57493 | 8.27295 | -0.01495 | 0.18104 |
| 63 | 9403 | $NiMn_2O_4$ | 8.4 | 8.40249 | -0.00249 | 0.02964 | 8.35788 | 0.04212 | 0.50143 |
| 64 | [36] | $NiRh_2O_4$ | 8.36 | 8.37154 | -0.01154 | 0.13804 | 8.34479 | 0.01521 | 0.18194 |
| 65 | 30076 | $PdZn_2O_4$ | 8.509 | 8.3175 | 0.1915 | 2.25052 | 8.3045 | 0.2045 | 2.40334 |
| 66 | 23498 | $RuCo_2O_4$ | 8.241 | 8.24339 | -0.00239 | 0.02894 | 8.24101 | -1E-5 | 0.00012 |
| 67 | 845 | $SiCo_2O_4$ | 8.14 | 8.28292 | -0.14292 | 1.75575 | 8.21121 | -0.07121 | 0.87482 |
| 68 | 36 | $SiFe_2O_4$ | 8.2340 | 8.02102 | 0.21298 | 2.58657 | 7.96566 | 0.26834 | 3.25893 |
| 69 | 86504 | $SiMg_2O_4^*$ | 8.069 | 8.28197 | -0.21297 | 2.63931 | 8.19799 | -0.12899 | 1.59859 |
| 70 | 8134 | $SiNi_2O_4$ | 8.044 | 8.13222 | -0.08822 | 1.09672 | 8.07119 | -0.02719 | 0.33802 |
| 71 | 167193 | $SiZn_2O_4^*$ | 8.0755 | 8.29615 | -0.22065 | 2.73236 | 8.21859 | -0.14309 | 1.7719 |
| 72 | 167815 | $SnMg_2O_4^*$ | 8.525 | 8.27546 | 0.24954 | 2.92714 | 8.19491 | 0.33009 | 3.87202 |
| 73 | 18186 | $TiFe_2O_4$ | 8.521 | 8.78449 | -0.26349 | 3.09221 | 8.74667 | -0.22567 | 2.6484 |
| 74 | 36 | $TiMg_2O_4$ | 8.4450 | 8.33602 | 0.10898 | 1.29047 | 8.28014 | 0.16486 | 1.95216 |
| 75 | 75377 | $TiMn_2O_4$ | 8.6806 | 8.50901 | 0.17159 | 1.97675 | 8.4468 | 0.2338 | 2.69336 |
| 76 | 36 | $TiZn_2O_4$ | 8.4870 | 8.37697 | 0.11003 | 1.29639 | 8.32572 | 0.16128 | 1.90032 |
| 77 | 2133 | $WNa_2O_4$ | 9.133 | 9.27231 | -0.13931 | 1.52534 | 9.16011 | -0.02711 | 0.29684 |
| 78 | 75629 | $ZnAl_2O_4$ | 8.0867 | 8.18965 | -0.10295 | 1.27308 | 8.15475 | -0.06805 | 0.84151 |
| 79 | 171889 | $ZnCr_2O_4$ | 8.3291 | 8.39801 | -0.06891 | 0.82733 | 8.35033 | -0.02123 | 0.25489 |
| 80 | 66128 | $ZnFe_2O_4$ | 8.4465 | 8.45864 | -0.01214 | 0.14372 | 8.4106 | 0.0359 | 0.42503 |
| 81 | 81105 | $ZnGa_2O_4$ | 8.3342 | 8.39403 | -0.05983 | 0.71792 | 8.34986 | -0.01566 | 0.1879 |
| 82 | 109298 | $ZnRh_2O_4$ | 8.54 | 8.4601 | 0.0799 | 0.93566 | 8.4217 | 0.1183 | 1.38525 |
| 83 | 36 | $ZnV_2O_4$ | 8.409 | 8.4684 | -0.0594 | 0.70641 | 8.41539 | -0.00639 | 0.07599 |
| 84 | 43025 | $CdAl_2S_4$ | 10.24 | 9.99663 | 0.24337 | 2.37667 | 10.02676 | 0.21324 | 2.08242 |
| 85 | 39415 | $CdCr_2S_4$ | 10.24 | 10.20499 | 0.03501 | 0.34191 | 10.24239 | -0.00239 | 0.02334 |
| 86 | 52798 | $CdDy_2S_4$ | 11.26 | 11.05093 | 0.20907 | 1.85678 | 11.25699 | 0.00301 | 0.02673 |
| 87 | 100518 | $CdEr_2S_4$ | 11.1 | 10.98972 | 0.11028 | 0.99349 | 11.18614 | -0.08614 | 0.77604 |
| 88 | 37405 | $CdHo_2S_4$ | 11.24 | 11.02033 | 0.21968 | 1.9544 | 11.22157 | 0.01843 | 0.16397 |
| 89 | 108215 | $CdIn_2S_4$ | 10.797 | 10.68631 | 0.11069 | 1.02517 | 10.73955 | 0.05745 | 0.53209 |
| 90 | 37410 | $CdLu_2S_4$ | 10.945 | 10.90862 | 0.03638 | 0.33236 | 11.09151 | -0.14651 | 1.3386 |
| 91 | 94994 | $CdSc_2S_4$ | 10.726 | 10.5877 | 0.1383 | 1.28944 | 10.72324 | 0.00276 | 0.02573 |
| 92 | 41111 | $CdTm_2S_4$ | 11.085 | 10.9618 | 0.1232 | 1.11144 | 11.15363 | -0.06863 | 0.61912 |
| 93 | 61697 | $CdY_2S_4$ | 11.216 | 11.01881 | 0.19719 | 1.75815 | 11.22208 | -0.00608 | 0.05421 |
| 94 | 41112 | $CdYb_2S_4$ | 11.055 | 10.94704 | 0.10797 | 0.97662 | 11.17004 | -0.11504 | 1.04062 |
| 95 | 24212 | $CoCo_2S_4$ | 9.4055 | 9.90407 | -0.49857 | 5.30079 | 9.77169 | -0.36619 | 3.89336 |
| 96 | 52942 | $CoCr_2S_4$ | 9.923 | 9.94291 | -0.01991 | 0.20067 | 9.86151 | 0.06149 | 0.61967 |
| 97 | 36 | $CoIn_2S_4$ | 10.559 | 10.42424 | 0.13476 | 1.2763 | 10.35866 | 0.20034 | 1.89734 |
| 98 | 42 | $CoNi_2S_4$ | 9.424 | 9.87383 | -0.44983 | 4.7732 | 9.73233 | -0.30833 | 3.27175 |
| 99 | 174043 | $CoRh_2S_4$ | 9.805 | 10.005 | -0.2 | 2.03976 | 9.79484 | 0.01016 | 0.10362 |
| 100 | 43527 | $CrAl_2S_4$ | 9.914 | 9.8313 | 0.0827 | 0.83419 | 9.85984 | 0.05416 | 0.5463 |
| 101 | 43528 | $CrIn_2S_4$ | 10.59 | 10.39325 | 0.19675 | 1.85793 | 10.36471 | 0.22529 | 2.12738 |
| 102 | 52942 | $CuCo_2S_4$ | 9.923 | 9.88982 | 0.03318 | 0.33433 | 9.74838 | 0.17462 | 1.75975 |
| 103 | 625675 | $CuCr_2S_4$ | 9.813 | 9.92867 | -0.11567 | 1.17844 | 9.8382 | -0.0252 | 0.25680 |
| 104 | 75531 | $CuIr_2S_4$ | 9.8474 | 10.04017 | -0.19277 | 1.95754 | 9.84254 | 0.00486 | 0.04935 |
| 105 | 41900 | $CuRh_2S_4$ | 9.788 | 9.99076 | -0.20275 | 2.07147 | 9.77153 | 0.01647 | 0.16827 |
| 106 | 170227 | $CuTi_2S_4$ | 10.0059 | 10.08978 | -0.08378 | 0.83733 | 10.03927 | -0.03327 | 0.3325 |
| 107 | 10035 | $CuV_2S_4$ | 9.8 | 9.99906 | -0.19906 | 2.03124 | 9.9212 | -0.1212 | 1.23673 |
| 108 | 27027 | $CuZr_2S_4$ | 10.378 | 10.24793 | 0.13007 | 1.25334 | 10.25658 | 0.12142 | 1.16997 |
| 109 | 95399 | $FeCr_2S_4$ | 9.9756 | 10.0087 | -0.0331 | 0.33183 | 9.95774 | 0.01786 | 0.17904 |
| 110 | 42535 | $FeFe_2S_4$ | 9.876 | 10.06933 | -0.19333 | 1.95759 | 9.98685 | -0.11085 | 1.12242 |
| 111 | 68411 | $FeIn_2S_4$ | 10.618 | 10.49003 | 0.12797 | 1.20526 | 10.4549 | 0.1631 | 1.53607 |
| 112 | 71678 | $FeLu_2S_4$ | 10.786 | 10.71234 | 0.07366 | 0.68296 | 10.80686 | -0.02086 | 0.1934 |
| 113 | 42590 | $FeNi_2S_4$ | 9.465 | 9.93962 | -0.47462 | 5.01443 | 9.82856 | -0.36356 | 3.8411 |
| 114 | 174045 | $FeRh_2S_4$ | 9.902 | 10.07079 | -0.16879 | 1.70458 | 9.89107 | 0.01093 | 0.11038 |
| 115 | 37425 | $FeSc_2S_4$ | 10.525 | 10.39141 | 0.13359 | 1.26927 | 10.43859 | 0.08641 | 0.821 |
| 116 | 37419 | $FeYb_2S_4$ | 10.838 | 10.75075 | 0.08725 | 0.80505 | 10.88539 | -0.04739 | 0.43726 |
| 117 | 608160 | $HgAl_2S_4$ | 10.28 | 10.18036 | 0.09964 | 0.9693 | 10.1744 | 0.1056 | 1.02724 |
| 118 | 53129 | $HgCr_2S_4$ | 10.235 | 10.38871 | -0.15372 | 1.50186 | 10.39003 | -0.15503 | 1.5147 |
| 119 | 56081 | $HgIn_2S_4$ | 10.812 | 10.87004 | -0.05804 | 0.5368 | 10.88719 | -0.07519 | 0.69543 |
| 120 | 53096 | $MgIn_2S_4$ | 10.687 | 10.47395 | 0.21305 | 1.99352 | 10.57474 | 0.11226 | 1.05044 |
| 121 | 37420 | $MgLu_2S_4$ | 10.949 | 10.69626 | 0.25274 | 2.30832 | 10.9267 | 0.0223 | 0.20367 |
| 122 | 37423 | $MgSc_2S_4$ | 10.627 | 10.37534 | 0.25167 | 2.36817 | 10.55843 | 0.06857 | 0.64524 |
| 123 | 37417 | $MgYb_2S_4$ | 10.957 | 10.73467 | 0.22233 | 2.02907 | 11.00523 | -0.04823 | 0.44018 |
| 124 | 53133 | $MnCr_2S_4$ | 10.110 | 10.07528 | 0.03472 | 0.34345 | 10.11692 | -0.00692 | 0.06845 |
| 125 | 65986 | $MnIn_2S_4$ | 10.72 | 10.5566 | 0.1634 | 1.52424 | 10.61407 | 0.10593 | 0.98815 |
| 126 | 37421 | $MnLu_2S_4$ | 10.921 | 10.77891 | 0.14209 | 1.30106 | 10.96603 | -0.04503 | 0.41232 |
| 127 | 37424 | $MnSc_2S_4$ | 10.623 | 10.45798 | 0.16502 | 1.55338 | 10.59777 | 0.02523 | 0.2375 |
| 128 | 37418 | $MnYb_2S_4$ | 10.949 | 10.81732 | 0.13168 | 1.20263 | 11.04456 | -0.09556 | 0.87277 |
| 129 | 23773 | $NiCo_2S_4$ | 9.424 | 9.86459 | -0.44059 | 4.67521 | 9.71395 | -0.28995 | 3.07672 |
| 130 | 53103 | $NiIn_2S_4$ | 10.505 | 10.38476 | 0.12024 | 1.14459 | 10.30092 | 0.20408 | 1.94269 |
| 131 | 36271 | $NiNi_2S_4$ | 9.457 | 9.83435 | -0.37735 | 3.99019 | 9.67459 | -0.21759 | 2.30084 |
| 132 | 105326 | $NiRh_2S_4$ | 9.6 | 9.96552 | -0.36552 | 3.80753 | 9.73709 | -0.13709 | 1.42802 |
| 133 | 53065 | $RhCo_2S_4$ | 9.67 | 9.80455 | -0.13455 | 1.3914 | 9.5525 | 0.1175 | 1.2151 |
| 134 | 53524 | $RhFe_2S_4$ | 9.87 | 9.94418 | -0.07418 | 0.75153 | 9.71507 | 0.15493 | 1.56971 |
| 135 | 105326 | $RhNi_2S_4$ | 9.6 | 9.72077 | -0.12077 | 1.25803 | 9.45495 | 0.14505 | 1.51094 |
| 136 | 35380 | $ZnAl_2S_4$ | 10.009 | 9.78363 | 0.22537 | 2.25164 | 9.76957 | 0.23943 | 2.39215 |
| 137 | 42019 | $ZnCr_2S_4$ | 9.982 | 9.99199 | -0.00999 | 0.1001 | 9.9852 | -0.0032 | 0.03206 |
| 138 | 81811 | $ZnIn_2S_4$ | 10.622 | 10.47332 | 0.14868 | 1.39977 | 10.48236 | 0.13964 | 1.31463 |
| 139 | 36 | $ZnSc_2S_4$ | 10.478 | 10.3747 | 0.1033 | 0.98588 | 10.46605 | 0.01195 | 0.11405 |
| 140 | 51423 | $CdAl_2Se_4^*$ | 10.73 | 10.5337 | 0.1963 | 1.82949 | 10.6894 | 0.0406 | 0.37838 |
| 141 | 78554 | $CdCr_2Se_4$ | 10.7346 | 10.74206 | -0.00706 | 0.06572 | 10.83234 | -0.09734 | 0.90675 |
| 142 | 246499 | $CdDy_2Se_4$ | 11.647 | 11.58799 | 0.05901 | 0.50662 | 11.58674 | 0.06026 | 0.51739 |
| 143 | 37406 | $CdEr_2Se_4$ | 11.603 | 11.52679 | 0.07621 | 0.65681 | 11.53536 | 0.06764 | 0.58295 |
| 144 | 40583 | $CdHo_2Se_4$ | 11.631 | 11.55739 | 0.07361 | 0.63286 | 11.56105 | 0.06995 | 0.60141 |
| 145 | 52811 | $CdIn_2Se_4$ | 11.345 | 11.22338 | 0.12162 | 1.07202 | 11.16131 | 0.18369 | 1.61913 |
| 146 | 620129 | $CdLu_2Se_4$ | 11.515 | 11.44569 | 0.06931 | 0.60191 | 11.46634 | 0.04866 | 0.42258 |
| 147 | 620411 | $CdSc_2Se_4$ | 11.208 | 11.12476 | 0.08324 | 0.74267 | 11.20096 | 0.00704 | 0.06281 |
| 148 | 40582 | $CdTm_2Se_4$ | 11.56 | 11.49886 | 0.06114 | 0.52886 | 11.51168 | 0.04832 | 0.41799 |
| 149 | 620457 | $CdY_2Se_4$ | 11.66 | 11.55587 | 0.10413 | 0.89304 | 11.56262 | 0.09738 | 0.83516 |
| 150 | 37408 | $CdYb_2Se_4$ | 11.528 | 11.4841 | 0.0439 | 0.38079 | 11.54121 | -0.01321 | 0.11459 |
| 151 | 42538 | $CoCo_2Se_4$ | 10.2 | 10.44113 | -0.24113 | 2.36405 | 10.42303 | -0.22303 | 2.18657 |
| 152 | 87477 | $CuCr_2Se_4$ | 10.3364 | 10.46574 | -0.12974 | 1.25519 | 10.49101 | -0.15501 | 1.49971 |
| 153 | 41903 | $CuRh_2Se_4$ | 10.264 | 10.52782 | -0.26382 | 2.57036 | 10.36978 | -0.10578 | 1.03059 |
| 154 | 608163 | $HgAl_2Se_4$ | 10.78 | 10.71742 | 0.06258 | 0.58049 | 10.76151 | 0.01849 | 0.17152 |
| 155 | 402408 | $HgCr_2Se_4$ | 10.7418 | 10.92578 | -0.18378 | 1.71087 | 10.90445 | -0.16245 | 1.51229 |
| 156 | 630754 | $MgEr_2Se_4$ | 11.475 | 11.31443 | 0.16057 | 1.39931 | 11.45949 | 0.01551 | 0.13516 |
| 157 | 44912 | $MgLu_2Se_4$ | 11.43 | 11.23333 | 0.19667 | 1.72066 | 11.39046 | 0.03954 | 0.34593 |
| 158 | 76051 | $MgTm_2Se_4$ | 11.469 | 11.2865 | 0.1825 | 1.59122 | 11.43581 | 0.03319 | 0.28939 |
| 159 | 76052 | $MgY_2Se_4$ | 11.57 | 11.34351 | 0.22649 | 1.95755 | 11.48674 | 0.08326 | 0.71962 |
| 160 | 76053 | $MgYb_2Se_4$ | 11.444 | 11.27174 | 0.17226 | 1.50523 | 11.46533 | -0.02133 | 0.18639 |
| 161 | 74407 | $MnSc_2Se_4$ | 11.106 | 10.99505 | 0.11095 | 0.999 | 11.1229 | -0.0169 | 0.15217 |
| 162 | 76225 | $MnYb_2Se_4$ | 11.42 | 11.35439 | 0.06561 | 0.57451 | 11.46314 | -0.04314 | 0.37776 |
| 163 | 609325 | $ZnAl_2Se_4$ | 10.61 | 10.3207 | 0.2893 | 2.72667 | 10.49582 | 0.11418 | 1.07615 |
| 164 | 150966 | $ZnCr_2Se_4$ | 10.46 | 10.52906 | -0.06906 | 0.66022 | 10.63875 | -0.17875 | 1.70889 |
| 165 | 71695 | $AgCr_2Te_4$ | 11.371 | 11.52062 | -0.14962 | 1.31581 | 11.10541 | 0.26559 | 2.33568 |
| 166 | 619806 | $CdDy_2Te_4$ | 11.38 | 12.3805 | -1.0005 | 8.79177 | 11.95604 | -0.57604 | 5.06186 |
| 167 | 43041 | $CuCr_2Te_4$ | 11.26 | 11.25825 | 0.00175 | 0.01559 | 10.86031 | 0.39969 | 3.54964 |
| 168 | [43] | $c\text{-}Si_3N_4^*$ | 7.8367 | 7.57818 | 0.25852 | 3.29881 | 7.7635 | 0.0732 | 0.93407 |
| 169 | [43] | $c\text{-}Ti_3N_4^*$ | 8.4459 | 8.4008 | 0.0451 | 0.53394 | 8.5555 | -0.1096 | 1.29767 |
| 170 | [44] | $c\text{-}C_3N_4^*$ | 6.8952 | 6.65724 | 0.23796 | 3.45108 | 6.9308 | -0.0356 | 0.5163 |
| 171 | [44] | $c\text{-}Ge_3N_4^*$ | 8.2110 | 8.05848 | 0.15252 | 1.85756 | 8.2578 | -0.0468 | 0.56997 |
| 172 | [44] | $c\text{-}Sn_3N_4^*$ | 8.9658 | 8.69116 | 0.27464 | 3.06322 | 8.8975 | 0.0683 | 0.76178 |
| 173 | [44] | $c\text{-}Zr_3N_4^*$ | 9.1217 | 8.96096 | 0.16074 | 1.76212 | 9.1393 | -0.0176 | 0.19295 |
| 174 | [44] | $c\text{-}CSi_2N_4^*$ | 7.5209 | 7.37491 | 0.14599 | 1.94118 | 7.4617 | 0.0592 | 0.78714 |
| 175 | [44] | $c\text{-}SiC_2N_4^*$ | 7.2867 | 6.86052 | 0.42618 | 5.84877 | 7.2326 | 0.0541 | 0.74245 |
| 176 | [44] | $c\text{-}CGe_2N_4^*$ | 7.7407 | 7.71016 | 0.03054 | 0.39451 | 7.7514 | -0.0107 | 0.13823 |
| 177 | [44] | $c\text{-}GeC_2N_4^*$ | 7.4284 | 7.00556 | 0.42284 | 5.69228 | 7.4373 | -0.0089 | 0.11981 |
| 178 | [44] | $c\text{-}SiGe_2N_4^*$ | 8.0871 | 7.91344 | 0.17366 | 2.1474 | 8.0531 | 0.034 | 0.42042 |
| 179 | [44] | $c\text{-}GeSi_2N_4^*$ | 8.0011 | 7.72322 | 0.27788 | 3.47302 | 7.9682 | 0.0329 | 0.41119 |
| 180 | [44] | $c\text{-}CTi_2N_4^*$ | 7.8351 | 7.96532 | -0.13022 | 1.66198 | 7.9161 | -0.081 | 1.03381 |
| 181 | [44] | $c\text{-}TiC_2N_4^*$ | 7.5400 | 7.09273 | 0.44727 | 5.932 | 7.5703 | -0.0303 | 0.40186 |
| 182 | 44 | c-$SiTi_2N_4^*$ | 8.2168 | 8.16859 | 0.0482 | 0.58666 | 8.2179 | -0.0011 | 0.01339 |
| 183 | 44 | c-$GeTi_2N_4^*$ | 8.4002 | 8.31363 | 0.08657 | 1.03055 | 8.4226 | -0.0224 | 0.26666 |
| 184 | 44 | c-$TiGe_2N_4^*$ | 8.3158 | 8.14565 | 0.17015 | 2.04614 | 8.3908 | -0.075 | 0.9019 |
| 185 | 44 | c-$TiZr_2N_4^*$ | 8.9276 | 8.73294 | 0.19466 | 2.1804 | 8.8103 | 0.1173 | 1.3139 |

Table 2. Predicted lattice parameters (all in Å) for the hypothetic $AM_2X_4$ ($A$=Zn, Cd, $M$=Sc, Y, Ln…Lu, $X$=O, S, Se) spinels

| | *A*=Zn, *X*=O, oxides | | *A*=Cd, *X*=S, sulfides | | *A*=Cd, *X*=Se, selenides | |
|---|---|---|---|---|---|---|
| | Experiment | Predicted | Experiment | Predicted | Experiment | Predicted |
| $ASc_2X_4$ | | 8.70955 | | 10.70124 | | 11.2119 |
| $AY_2X_4$ | | 9.1218 | | 11.21339 | | 11.59803 |
| $ALa_2X_4$ | | 9.47294 | | 11.649809 | | 11.92707 |
| $ACe_2X_4$ | | 9.41441 | | 11.577072 | | 11.87223 |
| $APr_2X_4$ | | 9.36182 | | 11.513863 | | 11.82448 |
| $ANd_2X_4$ | | 9.34293 | | 11.489424 | | 11.8061 |
| $APm_2X_4$ | | 9.30996 | | 11.454218 | | 11.77932 |
| $ASm_2X_4$ | 9.228[45*] | 9.27589 | | 11.404176 | | 11.7419 |
| $AEu_2X_4$ | 9.214[45*] | 9.24515 | | 11.360680 | | 11.70933 |
| $AGd_2X_4$ | | 9.22181 | | 11.333839 | | 11.689 |
| $ATb_2X_4$ | | 9.1903 | | 11.324744 | | 11.68091 |
| $ADy_2X_4$ | | 9.15292 | 11.26 | 11.249173 | 11.647 | 11.62513 |
| $AHo_2X_4$ | | 9.12366 | 11.24 | 11.212804 | 11.631 | 11.59771 |
| $AEr_2X_4$ | | 9.0944 | 11.1 | 11.176436 | 11.603 | 11.57029 |
| $ATm_2X_4$ | | 9.06773 | 11.085 | 11.143050 | 11.56 | 11.54513 |
| $AYb_2X_4$ | | 9.04768 | 11.055 | 11.160720 | 11.528 | 11.55671 |
| $ALu_2X_4$ | | 9.01698 | 10.945 | 11.079259 | 11.515 | 11.49706 |

[*] These experimental data were not included into the main fit (Eqs. (1) – (2)), since the conditions of the samples preparations could not be verified and checked.

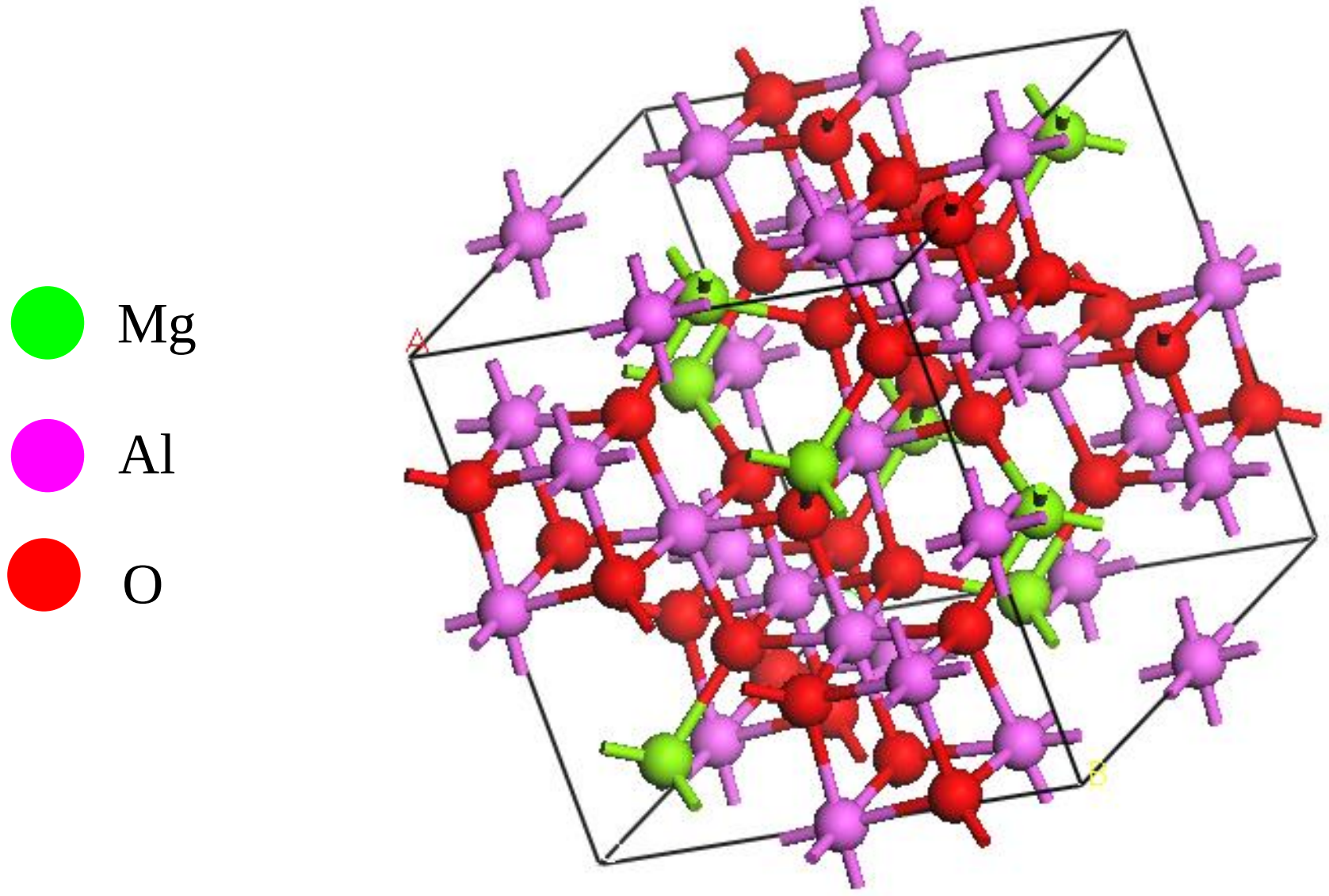

Fig. 1. (Colors online). One unit cell of $MgAl_2O_4$ as an example of the spinel's structure.

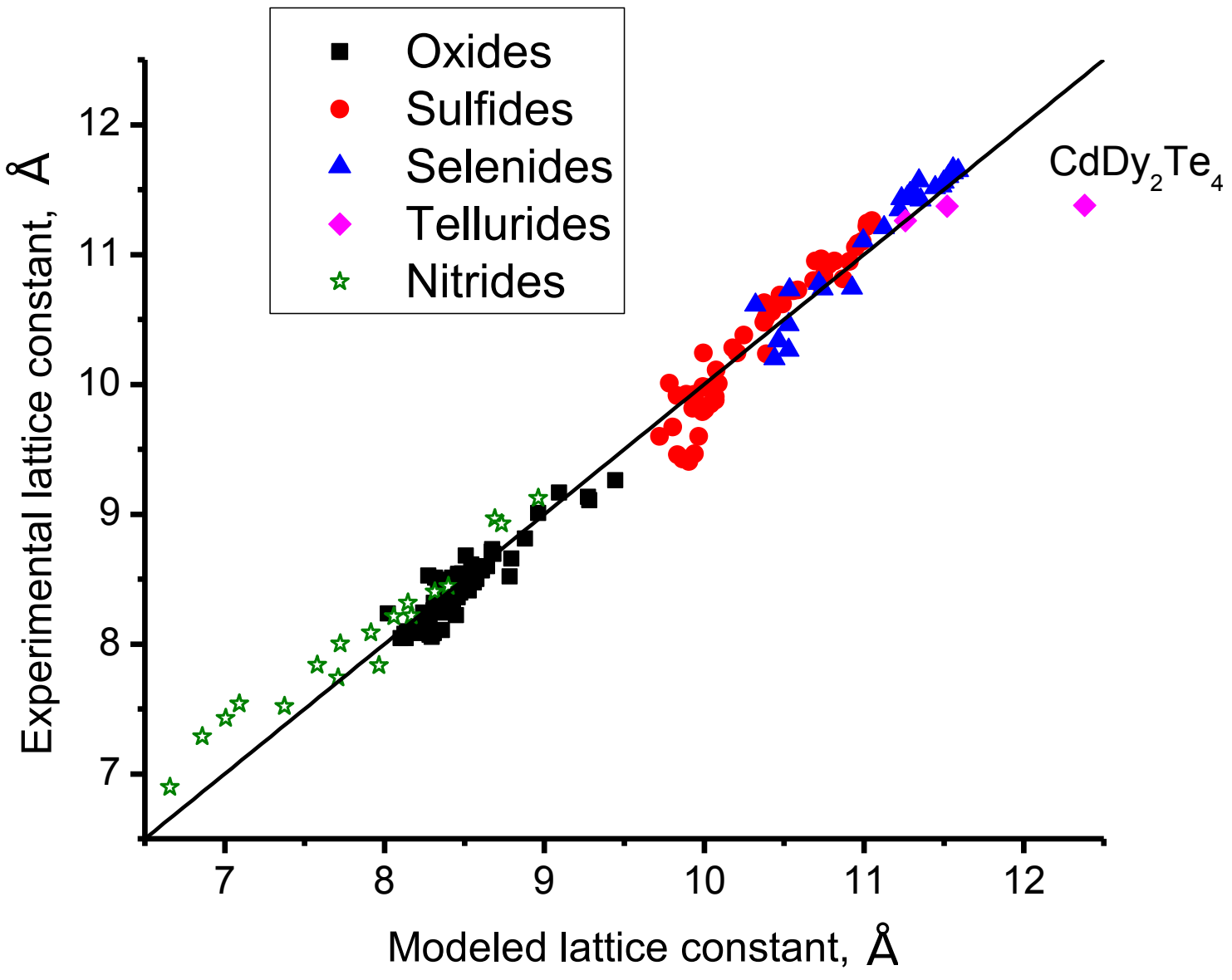

Fig. 2. (Colors online). Correlation between the calculated and experimental LCs in the group of 185 considered spinels as obtained by using Eq. (1).

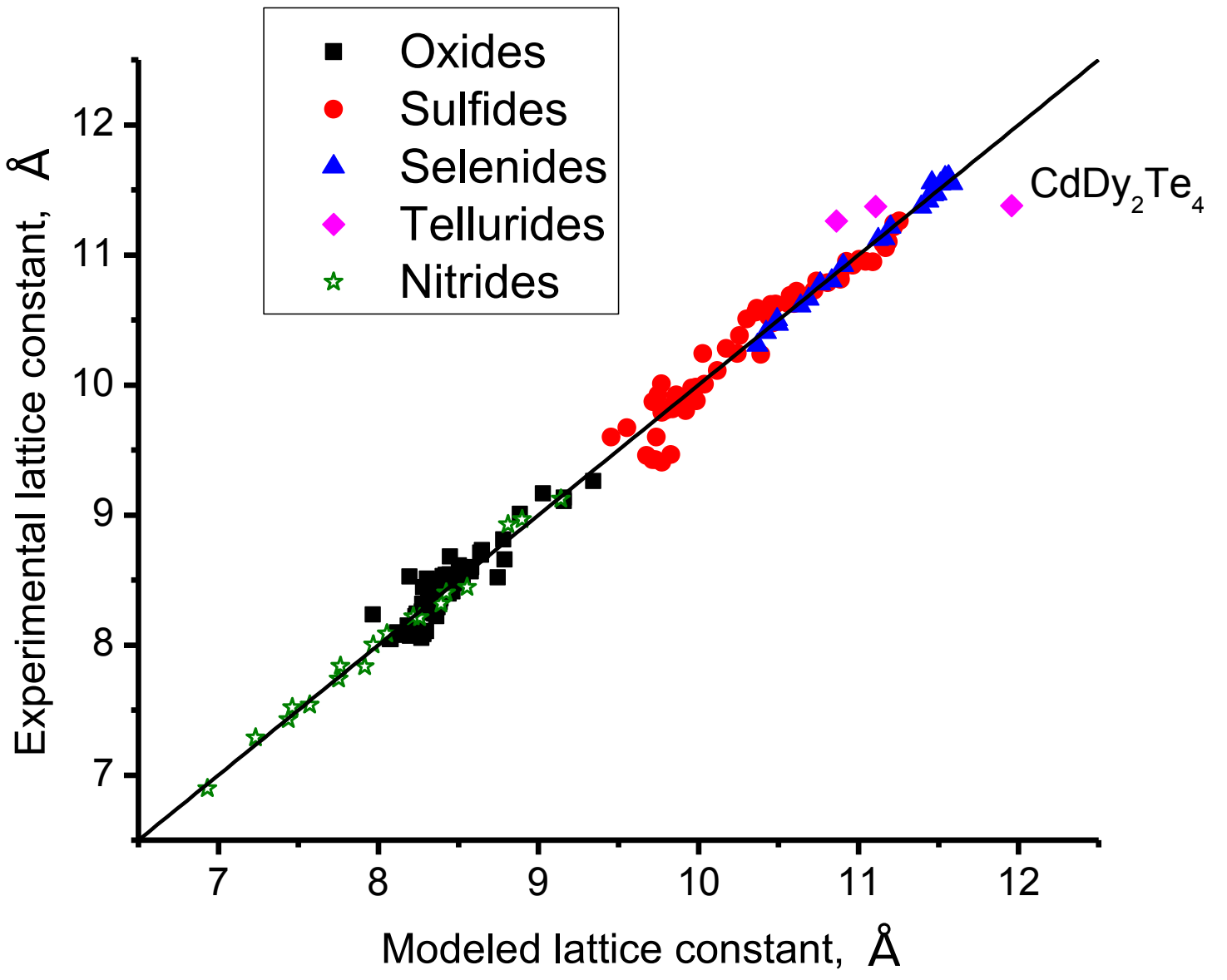

Fig. 3. (Colors online). Correlation between the calculated and experimental LCs in the group of 185 considered spinels as obtained by using Eqs. (2)-(5).

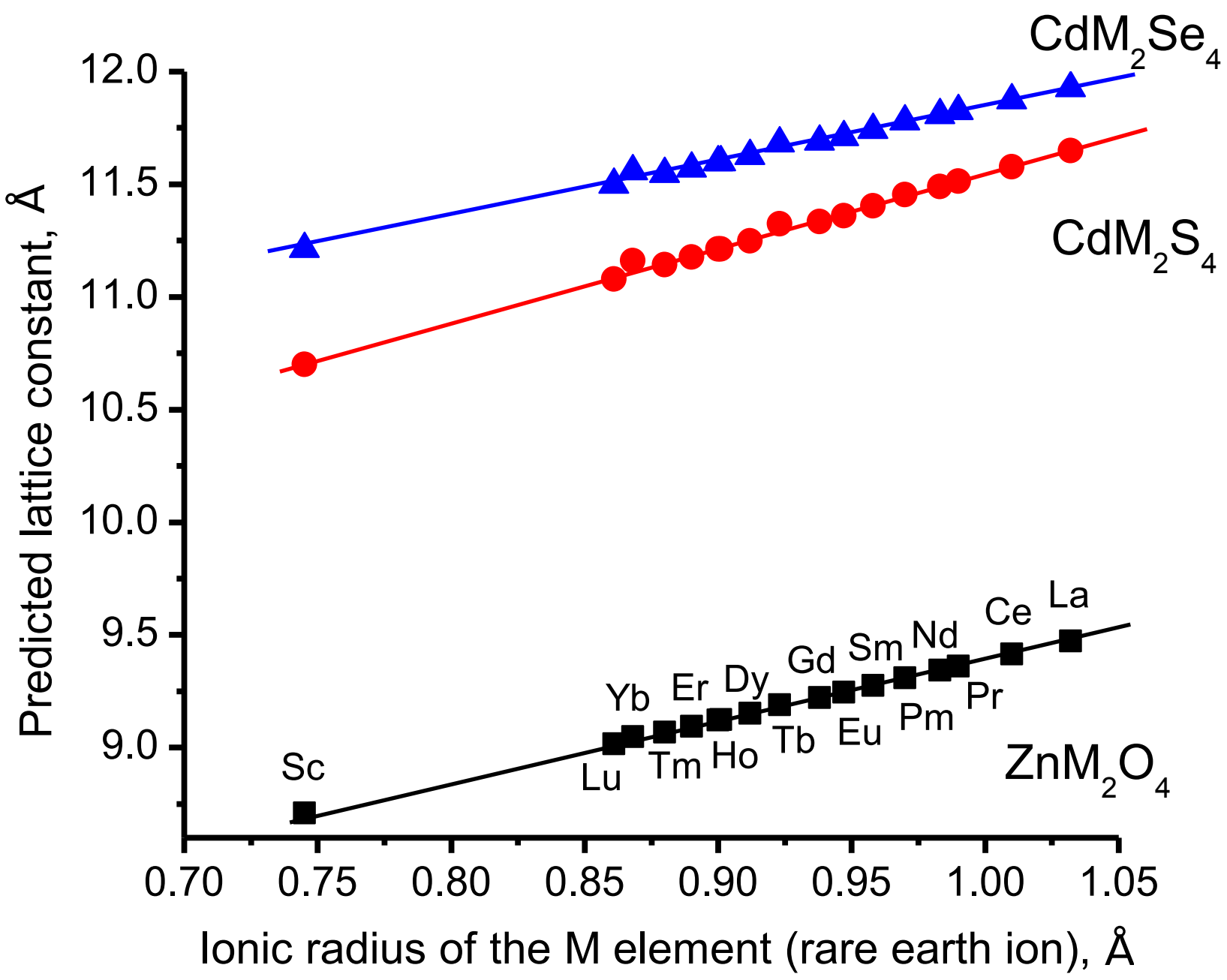

Fig. 4. Variation of the predicted lattice constants of the Zn$M_2O_4$, Cd$M_2S_4$ and Cd$M_2Se_4$ ($M$=Sc, Y, La-Lu) from Table 2 against the ionic radii of the rare earth ions. The straight lines are the guides to the eye only. The order of the data points in the two upper groups is the same as in the lowest one, where all $M$ ions are indicated.

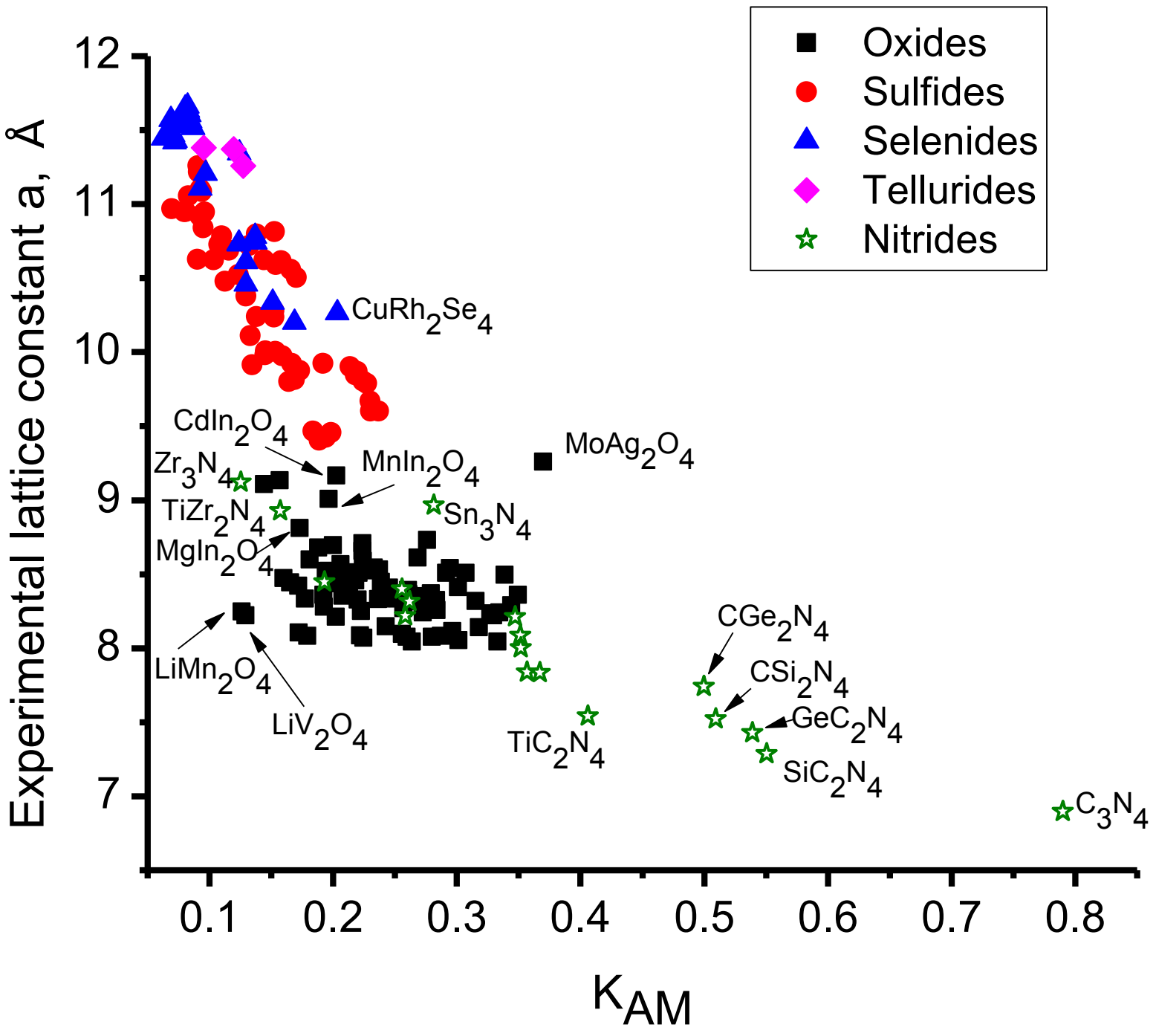

Fig. 5. (Colors online). Correlation between the experimental LCs and $K_{AM}$ value (Eq. (6)) in the group of 185 considered spinels.

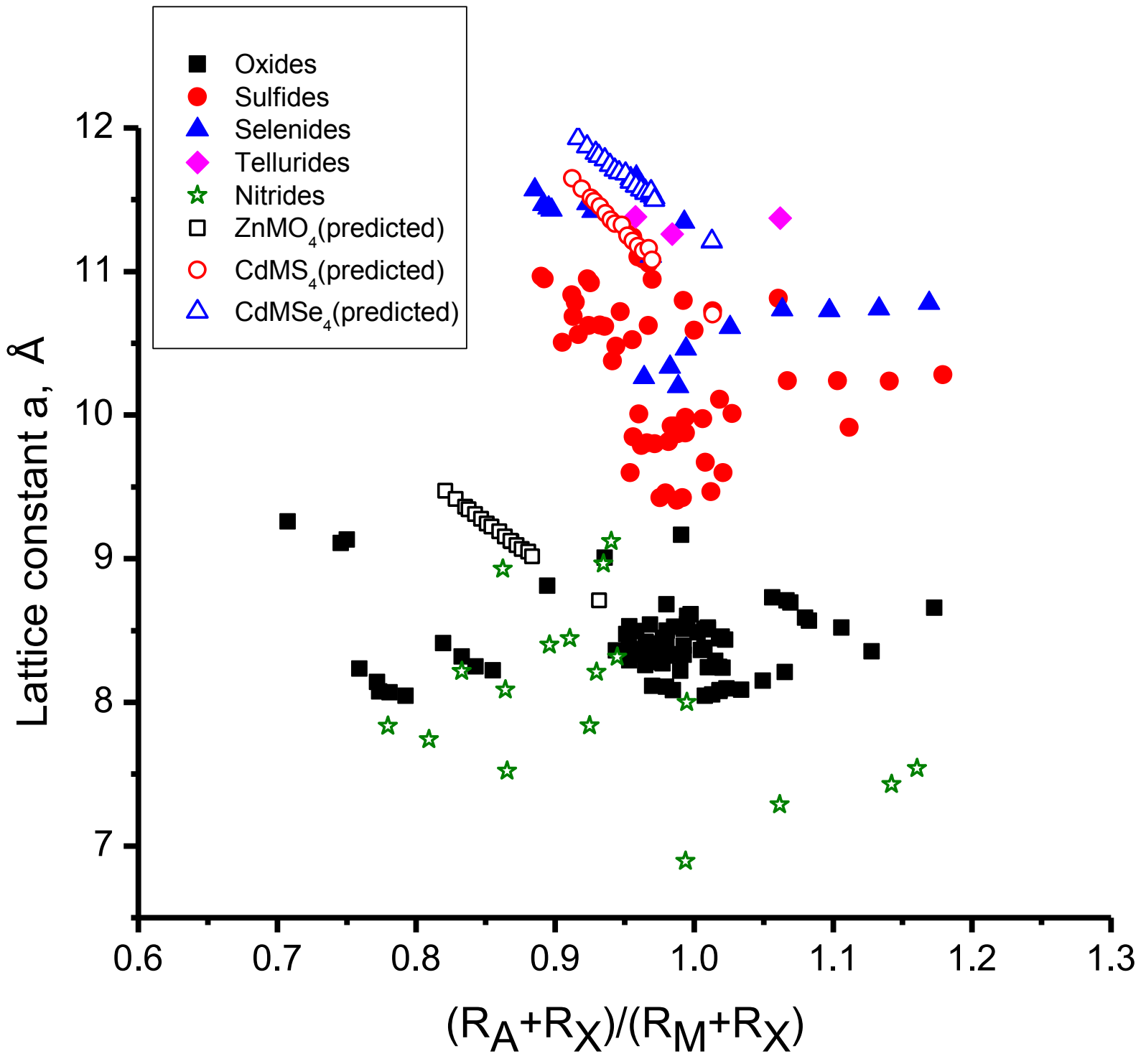

Fig. 6. (Colors online). Correlation between the experimental LCs and non-dimensional ratio of the sum of ionic radii $(R_A + R_X)/(R_M + R_X)$ in the group of 185 considered spinels. The predicted lattice constants of the Zn*M*$O_4$, Cd*M*$S_4$ and Cd*M*$Se_4$ (*M*=Sc, Y, La-Lu) from Table 2 are shown by the open squares, circles and triangles, respectively.

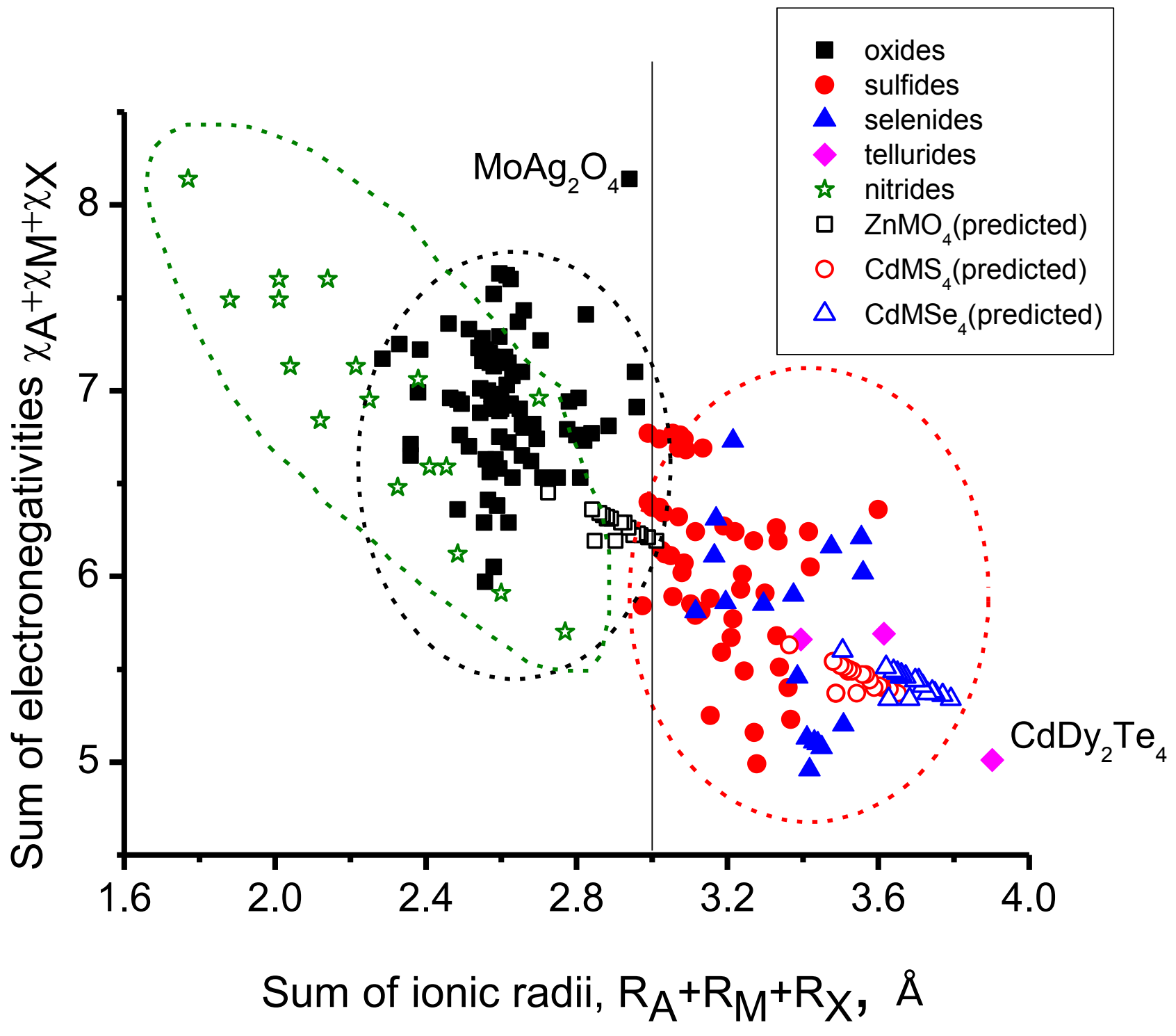

Fig. 7. (Colors online). Correlation between the sums of electronegativities and sum of ionic radii in the group of 185 considered spinels. The positions of the predicted spinels Zn*M*$O_4$, Cd*M*$S_4$ and Cd*M*$Se_4$ (*M*=Sc, Y, La-Lu) from Table 2 are shown by the open squares, circles and triangles, respectively.

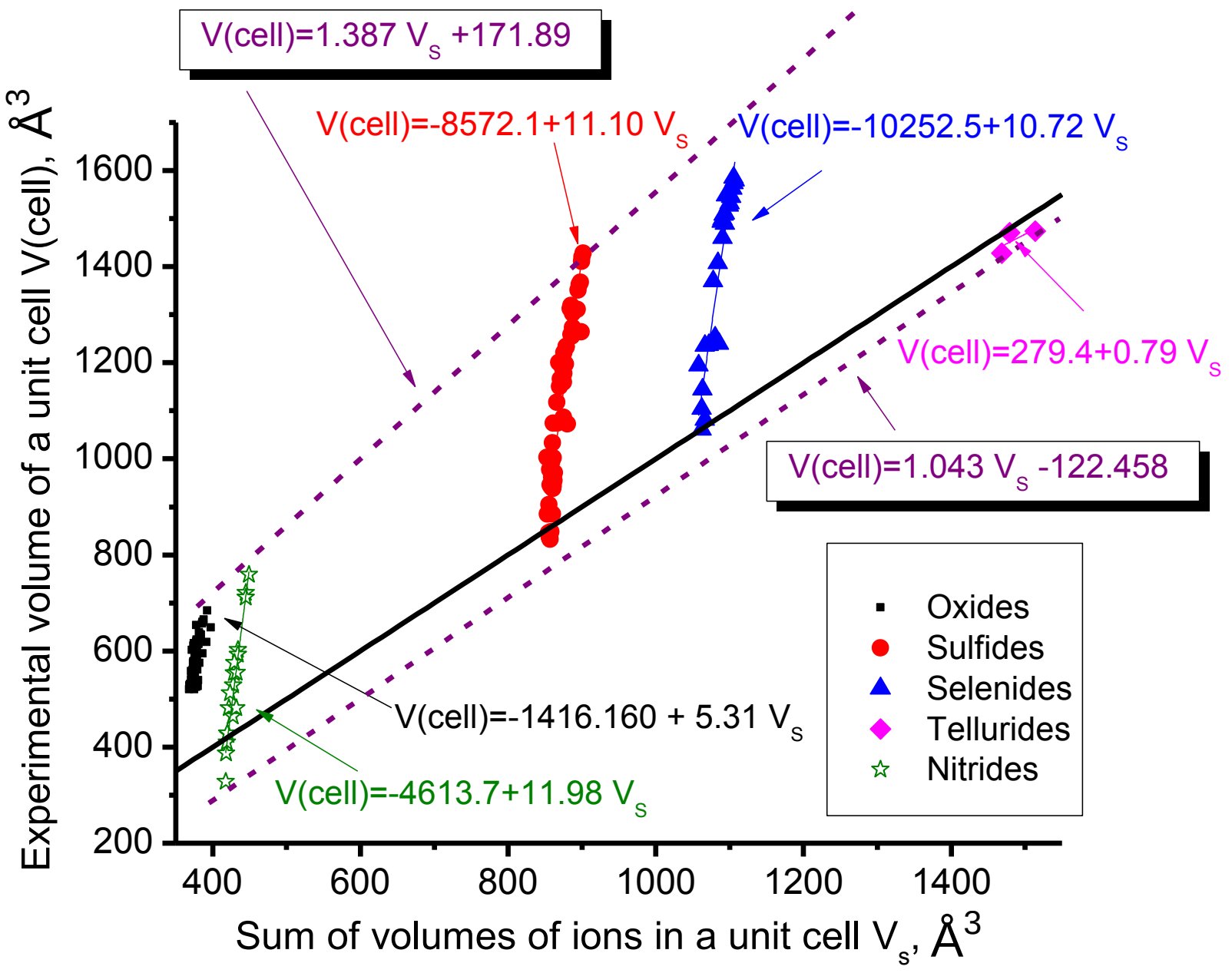

Fig. 8. (Colors online). Correlation between the experimental volume of the unit cell $V$(cell) and sum of volumes of ions $V_S$ in a unit cell in the group of 185 considered spinels. The black solid line corresponds to the condition $V$(cell)=$V_S$. See text for more details.

## References

(1) Pauling, L. The Nature of the Chemical Bond, 3rd ed., Cornell University Press, Ithaca, 1960.
(2) Martynov, A.I.; Batsanov, S.S. *Zh. Neorg. Khim*. **1980**, *25*, 3171-3175.
(3) Philips, J.C. *Phys. Rev. Lett*. **1968**, *20*, 550-553.
(4) Hinze, J.; Jaffe, H.H. *J. Am. Chem. Soc*. **1962**, *84*, 540-546.
(5) Allen, L.C. *J. Am. Chem. Soc*. **1992**, *114*, 1510-1511.
(6) Pauling, L. *J. Am. Chem. Soc*. **1932**, *54*, 3570-3582.
(7) Shannon, R.D. *Acta Cryst. A* **1976**, *32*, 751-767.
(8) Dimitrovska, S.; Aleksovska, S.; Kuzmanovski, I. *Centr. Eur. J. Chem*. **2005**, *3*, 198-215.
(9) Jiang, L.Q.; Guo, J.K.; Liu, H.B.; Zhou, M.; Zhou, X.; Wu, P.; Li, C.H. *J. Phys. Chem. Solids* **2006**, *67*, 1531-1536.
(10) Moreira, R.L.; Dias, A. *J. Phys. Chem. Solids* **2007**, *68*, 1617-1622.
(11) Verma, A.S.; Jindal, V.K. *J. Alloys Compds*. **2009**, *485*, 514-518.
(12) Majid, A.; Khan, A.; Javed, G.; Mirza, A.M. *Comput. Mater. Sci*. **2010**, *50*, 363-372.
(13) Brik, M.G.; Kityk, I.V. *J. Phys. Chem. Solids* **2011**, *72*, 1256-1260.
(14) Brik, M.G.; Srivastava, A.M. *J. Amer. Ceram. Soc*. **2012**, *95*, 1454-1460.
(15) Mouta, R.; Silva, R.X.; Paschoal, C.W.A. *Acta Cryst. B* **2013**, *69*, 439-445.
(16) Kuleshov, N.V.; Mikhailov, V.P.; Scherbitsky, V.G. *Proc. SPIE* **1994**, *2138*, 175-182.
(17) Jouini, A.; Yoshikawa, A.; Guyot, A.; Brenier, A.; Fukuda, T.; Boulon, G. *Opt. Mater*. **2007**, *30*, 47-49.
(18) Sanghera, J.; Bayya, S.; Villalobos, G.; Kim, W.; Frantz, J.; Shaw, B.; Sadowski, B.; Miklos, R.; Baker, C.; Hunt, M.; Aggarwal, I.; Kung, F.; Reicher, D.; Peplinski, S.; Ogloza, A.; Langston, P.; Lamar, C.; Varmette, P.; Dubinskiy, M.; DeSandre, L. *Opt. Mater*. **2011**, *33*, 511-518.
(19) Tshabalala, K.G.; Cho, S.H.; Park, J.K.; Pitale, S.S.; Nagpure, I.M.; Kroon, R.E.; Swart, H.C.; Ntwaeaborwa, O.M. *J. Alloys Compds*. **2011**, *509*, 10115-10120.
(20) Li, Y.X.; Niu, P.J.; Hu, L.; Xu, X.W.; Tang, C.C. *J. Lumin*. **2009**, *129*, 1204-1206.
(21) Deren, P.J.; Maleszka-Baginska, K.; Gluchowski, P.; Malecka, M.A. *J. Alloys Compds*. **2012**, *525*, 39-43.
(22) Sonoyama, N.; Kawamura, K.; Yamada, A.; Kanno, R. *J. Electrochem. Soc*. **2006**, *153*, H45-H50.
(23) Chen, Q.; Zhang, Z.J. *Appl. Phys. Lett.* **1998**, *73*, 3156-3158.
(24) Panda, R.N.; Gajbhiye, N.S.; Balaji, G. *J. Alloys Compds.* **2001**, *326*, 50-53.
(25) Falkovskaya, L.; Fishman, A.; Mitrofanov, V.; Tsukerblat, B. *Phys. Lett. A*, **2010**, *30*, 3067-3075.
(26) Satoh, T.; Tsushima, T.; Kudo, K. *Mater. Res. Bull*. **1974**, *9*, 1297-1300.
(27) O'Neill, H.St.C.; Navrotsky, A. *Amer. Mineral*. **1983**, *68*, 181-194.
(28) Price, G.D.; Price, S.L.; Burdett, J.K. *Phys. Chem. Minerals* **1982**, *8*, 69-76.
(29) Price, G.D. *Phys. Chem. Minerals* **1983**, *10*, 77-83.
(30) Ottonello, G. *Phys. Chem. Minerals* **1986**, *13*, 79-90.
(31) Burdett, J.K.; Price, G.D.; Price, S.L. *Phys. Rev. B* **1981**, *24*, 2903-2912.
(32) Barth, T.F.W.; Posnjak, E. *Z. Krist*. **1932**, *82*, 325-341.
(33) Sickafus, K.E.; Wills, J.M. *J. Am. Ceram. Soc*. **1999**, *82*, 3279-3292.
(34) Krok-Kowalski, J.; Warczewski, J.; Nikiforov, K. *J. Alloys Compds.* **2001**, *315*, 62-67.
(35) Kugimiya, K.; Steinfink, H. *Inorg. Chem*. **1968**, *7*, 1762-1770.
(36) Hill, R.J.; Craig, J.R.; Gibbs, G.V. *Phys. Chem. Minerals* **1979**, *4*, 317-339.
(37) Müller, U. *Inorganic Structural Chemistry Wiley & Sons*, **1993**.
(38) Lide, D.R. (Ed.) *Handbook of Chemistry and Physics*, CRC Press, **2004-2005**.
(39) http://www.fiz-karlsruhe.de/icsd.html
(40) Otero Arean, C.; Rodriguez Martinez, M.L.; Marta Arjona, A. *Mater. Chem. Phys*. **1983**, *8*, 443-450.
(41) Wessels, A.L.; Czekalla, R.; Jeitschko, W. *Mater. Res. Bull*. **1998**, *33*, 95-101.
(42) Huang, C.H.; Knop, O. *Canad. J. Chem*. **1971**, *49*, 598-602.
(43) Ching, W.Y.; Mo, S.-D.; Ouyang, L.; Tanaka, I.; Yoshiya, M. *Phys. Rev. B* **2000**, *61*, 10609-10614.
(44) Ching, W.Y.; Mo, S.-D.; Tanaka, I.; Yoshiya, M. *Phys. Rev. B* **2001**, *63*, 064102.
(45) Shaplygin, I.S.; Lazarev, V.B. *Zh. Neorg. Khimii* **1985**, *30*, 1595-1597.